# Trace element distribution in coarse-grained Ca-Al-rich inclusions by femtosecond LA-ICP-MS: Implications for U-Pb chronology

Merel E. Onderwater[a], Elishevah M. M. E. van Kooten[a], Glenn J. MacPherson[b], Alexander N. Krot[c], Martin Bizzarro[a,d]

[a]Centre for Star and Planet Formation, Globe Institute, University of Copenhagen, Copenhagen Denmark.
[b]Department of Mineral Sciences, National Museum of Natural History, Smithsonian Institution, Washington DC, USA.
[c]Hawai‘i Institute of Geophysics and Planetology, University of Hawai‘i at Mãnoa, Honolulu, HI, USA.
[d]Institut de Physique du Globe de Paris, Université de Paris, Paris, France.

Corresponding author: M. E. Onderwater (merel.onderwater@sund.ku.dk)

**Abstract**

Calcium-aluminium-rich inclusions (CAIs) in chondritic meteorites are the first solids that formed in the Solar System, as established by U-Pb chronology. After initial condensation, coarse-grained CAIs (Compact Type A, Type B, Type C) experienced melting, resulting in igneous partitioning and spatial redistribution of trace elements among constituent minerals. Understanding these distributions is essential for identifying trace-element host phases and interpreting U-Pb ages. Rare earth elements (REEs) and the actinides Th and U are particularly informative tracers, yet actinide behaviour in coarse-grained CAIs remains poorly constrained. Further, the decay of short-lived $^{247}$Cm to $^{235}$U ($t_{1/2}$ = 15.6 Myr) complicates U-Pb age determination of bulk CAIs, but its impact at the mineral scale remains elusive.

Here, we present major- and trace-element distribution maps of six coarse-grained CAIs mainly composed of melilite, Al,Ti-pyroxene, anorthite and spinel (±forsterite) from CV3 (Vigarano-type) chondrites acquired by femtosecond laser ablation ICP-MS. In CAIs lacking perovskite or hibonite, our results show that Al,Ti-pyroxene is the dominant host of U, Th, and most REEs, with strong enrichments observed toward pyroxene grain edges, whereas light REEs reside in melilite. Trace-element zonation reflects igneous fractional crystallization and is overprinted by melt evaporation, which caused depletion of the most volatile REEs (Yb, Eu, Ce) and U in CAI outer regions and rims. Using Nd and Th distribution maps as proxies for Cm, we model the effects of $^{247}$Cm-$^{235}$U decay on mineral-scale U isotopic compositions. Variable Cm-U partitioning among coexisting minerals is predicted to generate resolvable inter-mineral $^{235}$U/$^{238}$U variations, with the largest excesses in melilite. These variations can produce apparent U-Pb age offsets of ~0.1-1.0 Myr when U isotopic compositions and radiogenic Pb are derived from isotopically distinct mineral reservoirs, but generally remain small (< 0.15 Myr) when using bulk U isotopic compositions. Our results demonstrate that fractional crystallization and melt evaporation exert a primary control on U distribution and isotope systematics in igneous CAIs and provide new constraints on the interpretation and robustness of high-precision U-Pb chronology of the earliest Solar System solids.

## 1. Introduction

Calcium-aluminium-rich inclusions (CAIs) preserved in chondritic meteorites represent the oldest dated solids formed in the Solar System, as established by high-precision U-Pb chronology (Jacobsen et al., 2008; Amelin et al., 2010; Connelly et al., 2012). Their chemical and isotopic compositions provide fundamental constraints on the timing, conditions, and processes operating during the earliest stages of Solar System evolution, including evaporation, condensation, and melting (in the disk region near the protosun), and subsequent transport within the protoplanetary disk (Yoneda & Grossman, 1995; Krot et al., 2009; Ciesla, 2010; MacPherson, 2014).

CAIs are divided into two groups based on the size of mineral phases within them: fine-grained and coarse-grained CAIs. Coarse-grained igneous inclusions are thought to have formed by initial condensation from a nebular gas of near-solar composition (Grossman, 1973; Yoneda & Grossman, 1995), followed by one or more high-temperature melting events that produced their igneous textures and characteristic mineral assemblages, mainly consisting of melilite, Al,Ti-pyroxene, anorthite, spinel (±forsterite). Flash-melting episodes are thought to produce the characteristic Wark-Lovering (WL) rims around these inclusions (Wark & Lovering, 1977; MacPherson, 2014; MacPherson et al., 2026). As a result, coarse-grained CAIs archive a multi-stage formation history in which nebular signatures may be partially preserved but subsequently modified by igneous crystallization and melt evaporation.

Trace elements, particularly the rare earth elements (REEs) and the actinides Th and U, are sensitive tracers of both nebular and igneous processes. Their relative abundances and distributions among CAI minerals record volatility-controlled fractionation during condensation as well as mineral-melt partitioning during crystallization (Mason & Martin, 1977; Beckett et al., 1990; Hu et al., 2021). Despite their importance, actinides are often excluded from *in situ* studies because of their low (sub-parts per million) concentrations in CAI minerals. Consequently, the mineral hosts of U and Th in coarse-grained CAIs remain poorly constrained.

This knowledge gap directly impacts the interpretation of U-Pb ages of CAIs. The canonical absolute age of CAI formation (~4567.3 Ma) is defined by U-corrected Pb-Pb dating of only a small number of inclusions (n = 4; Amelin et al., 2010; Connelly et al., 2012). Discrepancies between the U-Pb and other chronometers have recently reopened discussion on the robustness of this canonical age (Desch et al., 2023; Piralla et al., 2023). Interpreting U-Pb ages requires understanding how U is distributed within CAIs and which minerals host the majority of the U budget (Amelin & Yin, 2025). Minor phases such as perovskite may host significant U (Stapanian, 1981; Murrell & Burnett, 1987; Yi et al., 2024), yet this phase is absent from many CAIs (particularly from coarse-grained CAIs), leaving the principal U host unresolved.

An additional complication arises from U isotopic variability in CAIs, which is at least partially caused by the decay of the now-extinct radionuclide $^{247}$Cm to $^{235}$U with a half-life of 15.6 Myr (Brennecka et al., 2010; Tissot et al., 2016). Bulk CAIs show resolvable variations in $^{235}$U/$^{238}$U that must be accounted for in Pb-Pb age calculations. Implicit in most chronological applications, however, is the assumption that U isotopic compositions are homogeneous within individual CAIs. Emerging evidence from large, coarse-grained inclusions suggests that this assumption may not always hold (Li, 2024), raising the possibility of mineral-scale U isotopic heterogeneity and associated age offsets for which the underlying cause remains unclear.

In this study, we address these issues by presenting the first comprehensive trace-element distribution analysis of coarse-grained CAIs acquired by femtosecond laser ablation ICP-MS. We quantify the abundances and spatial distributions of REEs, Th, and U across major CAI minerals and textural zones, with the goal of identifying the principal actinide host phases and evaluating the effects of igneous partitioning and melt evaporation. Using the measured Nd and Th abundances as proxies for Cm, we further explore the extent to which variable Cm-U partitioning among coexisting minerals can generate inter-mineral U isotopic variability and influence U-Pb chronometry. Our results provide new constraints on the internal geochemical architecture of igneous CAIs and its implications for high-precision chronology of the earliest Solar System solids.

## 2. Methods

### *2.1 Sample characterization*

Six large, coarse-grained CAIs were selected for this study because their size makes them potentially suitable for future U-Pb chronological studies. Two inclusions from the CV3 chondrite Leoville (CAIs L1B-1, L1B-2) were extracted from a slab of their host chondrite using a 250 μm diamond-coated wire saw. Four inclusions from the recently classified CV3 chondrite Northwest Africa (NWA) 18679 (CAIs #30, #44, #54, #55) were extracted from slices of their host chondrite (specimen USNM 7614) via biscuit cutting at the Smithsonian Institution.

Both host chondrites belong to the reduced subgroup of CV chondrites and have experienced only limited parent-body processing, making them well suited for the investigation of primary CAI characteristics. Leoville is one of the least altered CV chondrites in the meteorite record and exhibits a low degree of thermal metamorphism (petrologic type 3.1-3.2; Bonal et al., 2016) together with weak shock effects (shock stage S3; Scott et al., 1992). A petrographic investigation of NWA 18679 indicates very limited shock effects (S1-S2) based on undulatory extinction in olivine but the absence of mosaicism. This host chondrite has undergone extensive terrestrial weathering to the extent that no primary fusion crust is preserved. Iron oxides and calcite veins crosscutting NWA 18679, and elevated Cl abundances are most likely of terrestrial origin.

At the University of Copenhagen, the six selected CAIs were mounted in epoxy and polished with Milli-Q water using $Al_2O_3$ polishing grids. They were characterized using high-resolution backscatter electron (BSE) imaging and energy dispersive X-ray (EDS) mapping with a Zeiss EVO 15 scanning electron microscope (SEM) equipped with a Bruker X-Trace EDS detector at the Centre for Star and Planet Formation, University of Copenhagen. High-resolution BSE and EDS images of the selected inclusions and their rims were acquired in variable pressure mode at $2\times10^{-1}$ mbar, 20 kV accelerating voltage and a beam current ranging between 1.0 and 5.0 nA. Energy dispersive X-ray maps were acquired for Mg, Ca, Al, Ti, Si, Fe, Ni, Na, S, Cl, Cr, K and P.

Additionally, the NWA 18679 CAIs (#30, #44, #54, #55) were characterized by BSE and full spectrum imaging at the Smithsonian Institution using a Thermo Fisher FEI Quattro Analytical SEM. Major and minor compositions of minerals were analyzed using a JEOL JXA-8500F field-emission electron microprobe, operated at 15 kV accelerating voltage, 15 nA beam current and 1 μm beam size, using five wavelength spectrometers, also located at the Smithsonian Institution. Natural and synthetic minerals were used as standards. Data were corrected using a proprietary JEOL ZAF program.

*2.2 Elemental mapping by femtosecond LA-ICP-MS*

Major- and trace-element concentrations of the inclusions were measured by femtosecond laser ablation inductively coupled plasma mass spectrometry (fs-LA-ICP-MS) using the ESI NWRFemto laser ablation system connected to the Thermo Scientific iCAP RQ ICP-MS, located at the Centre for Star and Planet Formation, University of Copenhagen. The regions targeted by fs-LA-ICP-MS mapping covered a representative mineralogy of the inclusions, avoided large fractures and extended from inclusion core to rim. In addition, sections including the CAI rim and adjacent chondrite matrix were sampled in four of the six studied inclusions.

Before the start of measurements, the $N_2$ gas flow rate was stabilized at 1500 ml/min to remove atmospheric gases from the laser chamber, which took approximately 15 minutes for each session. The laser was heated for at least 10 minutes for a stable laser output energy. Prior to data acquisition, the signal intensity of the ICP-MS was optimized by tuning on the NIST 612 glass standard, and gas flow rates were adapted to maintain low oxide production rates by monitoring $^{232}Th^{16}O/^{232}Th^{+}$ (typically <0.4 %). During tuning, the $^{232}Th/^{238}U$ ratio was maintained at $1.0 \pm 0.03$. Helium was used as the carrier gas and had a flow rate of 800 ml/min.

Data were acquired with a laser fluence of 6-7 J/cm$^2$, a spot size of 20 μm, at a scan speed of 40 μm/s and 100 Hz repetition rate. The elemental maps were obtained in line scan mode, where we applied a 50% overlap on each line, equal to 10 μm, to increase our spatial resolution. This strategy significantly improves sampling density and signal continuity along mapped profiles. The following isotopes were analysed during all sessions: $^{23}$Na, $^{24}$Mg, $^{27}$Al, $^{29}$Si, $^{44}$Ca, $^{49}$Ti, $^{51}$V, $^{52}$Cr, $^{57}$Fe, $^{60}$Ni, $^{88}$Sr, $^{90}$Zr, $^{95}$Mo, $^{139}$La, $^{140}$Ce, $^{141}$Pr, $^{146}$Nd, $^{147}$Sm, $^{153}$Eu, $^{157}$Gd, $^{159}$Tb, $^{163}$Dy, $^{165}$Ho, $^{166}$Er, $^{169}$Tm, $^{172}$Yb, $^{175}$Lu, $^{208}$Pb, $^{232}$Th and $^{238}$U. During the session in which Leoville inclusions were mapped, additional isotopes $^{89}$Y, $^{45}$Sc, $^{135}$Ba, $^{137}$Ba and $^{151}$Eu were measured. The latter three additions were due to a suspected BaO interference on the $^{151}$Eu and $^{153}$Eu masses, which proved negligible. Each sample was bracketed by a standard block consisting of NIST 610, NIST 612 and BIR-1G glass standards and a BHVO-2 pressed powder pellet. We ensured that the duration of each sample run was less than three hours, to correct for signal drift during our sessions.

Data reduction was performed offline through the Iolite v.4 software and the 3D trace element data reduction scheme (Paton et al., 2011; Paul et al., 2012; Petrus et al., 2017). Yields of external standards were typically normalized to BHVO-2, and the standards were further internally normalized to their Al content. Our data quality can be assessed by the deviation of measured values from the accepted elemental abundances of our standards. Figure 1 shows the deviations of BIR-1G in various sessions, in which all elements are within acceptable values (< 20%) except for Mo, an element non-essential to our data analysis. Because Al concentrations vary among CAI minerals, the samples require a different internal normalization approach than normalization to Al content. To address this, a criterion was defined for each distinguishable phase within the targeted region, based on the signal intensity of characteristic elements in that phase (e.g. Ti for Al,Ti-pyroxene; Paul et al., 2023). These internal standard criteria normalized the total concentration of all measured elements to 100% for the corresponding phases. The regions of the sample that correspond to this phase criterion are normalized in the same manner, effectively minimizing matrix effects (Paul et al., 2023).

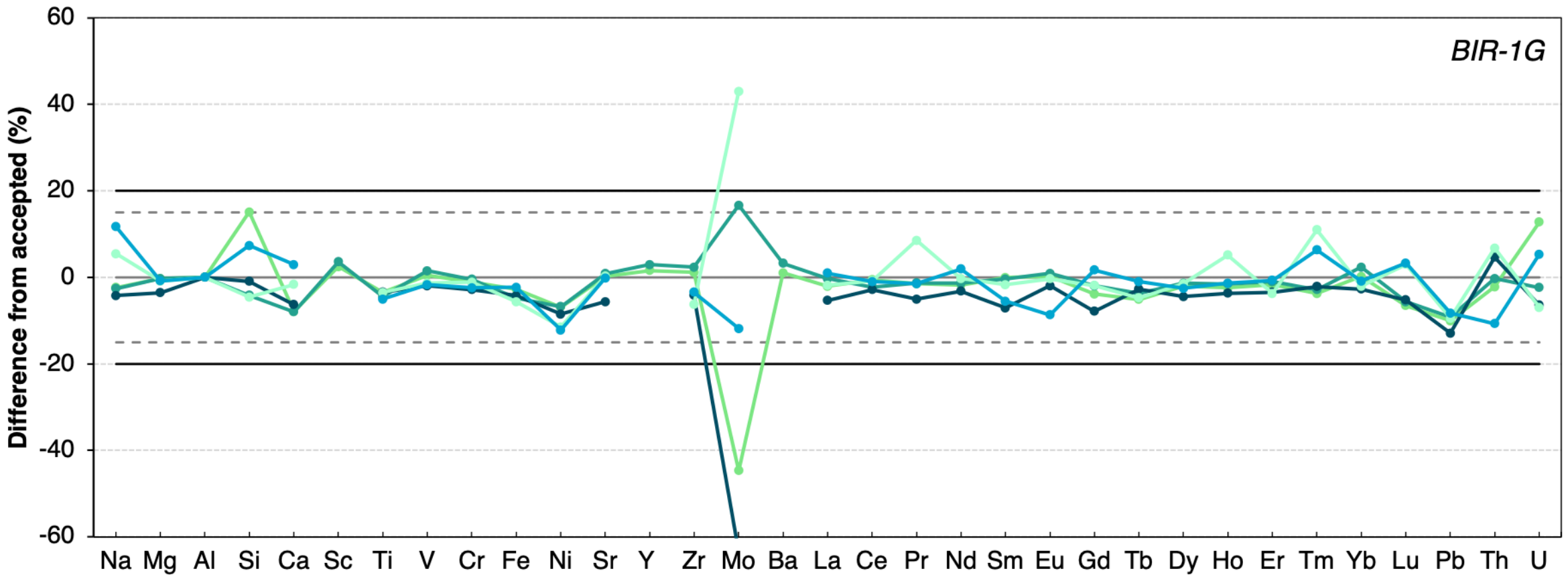

**Figure 1**: fs-LA-ICP-MS data quality of external standard BIR-1G
Difference (%) between measured and accepted elemental abundances in BIR-1G glass standard using fs-LA-ICP-MS, calculated as a secondary check in Iolite v.4 to assess the data quality. Each colour represents a different measurement session used for the calibration curves. The measured concentrations of all elements are consistently within 20% deviation from accepted values for BIR-1G, except for Mo.

The coarse-grained texture of the studied CAIs enables identification of distinct phases and individual grains at our spatial resolution. To distinguish between phases and report mineral-specific measurements, regions of interest (ROIs) on the fs-LA-ICP-MS maps are selected based on two approaches: (1) selection by the internal standard criteria and (2) selection by comparison with BSE-EDS maps. If the mineral grain size was smaller than the laser spot size (20 μm), ROIs were only defined where the mineral has a high grain concentration, with more than ~90% of the ROI represented by this phase. Fine-grained minerals (~10 μm or less) in CAIs, such as spinel and perovskite, fell under this definition. In only three inclusions, spinel occurs in nodules or palisades where the spinel grain concentration was high enough to define ROIs. Larger phases frequently contain small spinel inclusions, resulting in the corresponding ROIs to reflect inter-mineral mixtures. Where possible, spinel grains were excluded from ROIs to minimize compositional mixing. To investigate core-to-rim relations, we further construct ROIs to define the CAI core, the outer 200-300 μm layer or outer melilite mantle, and the CAI rim.

Analytical sensitivity was sufficient to quantify trace-element abundances at ppb to tens-of-ppb levels across all major CAI minerals. Uranium and Lu represent the lowest-concentration elements measured (i.e as low as 3-6 ppb in anorthite) and have estimated detection limits of ~1-3 ppb (Pettke et al., 2012). Although some analyses approach these limits in U-poor phases (Yi et al., 2024), concentrations below this detection limit are uncommon in coarse-grained igneous CAIs. The use of BHVO-2 and BIR-1G as external reference materials is appropriate given the overlap between their trace-element abundances and those measured in CAI minerals.

*2.3 $^{247}$Cm-$^{235}$U decay modelling*

The radiogenic production of $^{235}$U through $^{247}$Cm decay affected the U isotopic composition of early formed solids in the Solar System (Tissot et al., 2016), especially if extensive Cm-U fractionation occurred. As no long-lived Cm isotopes exist today, the extent of $^{247}$Cm decay on the bulk U composition has been modelled with proxies for Cm such as Nd, Sm and Th (Brennecka et al., 2010; Tissot et al., 2016). Neodymium and Th have primarily been used to predict the geochemical behaviour of Cm during condensation, owing to their similar volatilities and ionic radii (Boynton, 1978; Brennecka et al., 2010). Fine-grained CAIs represent direct products of condensation, but this study focuses on coarse-grained igneous CAIs that are thought to crystallize from a melt (MacPherson, 2014). Boynton (1978) argued that light REEs, in particular Nd and Sm, can also act as proxies for Cm during igneous processes. While these elements provide a provisional basis for tracing Cm, the most appropriate proxy remains uncertain.

Given current analytical limitations, it is not yet feasible to measure mineral-specific U isotopic compositions in CAIs that are small (< 1 mm) and/or U-poor. Therefore, we explore through modelling if variable Cm-U partitioning can cause U isotopic variability between individual mineral phases in a single CAI, and how it would affect U-Pb chronometry of that CAI. Here, we approximate the behaviour of Cm with Nd and Th as proxies (Brennecka et al., 2010; Tissot et al., 2016). The U, Nd and Th concentrations measured by fs-LA-ICP-MS are used to calculate the atomic $^{144}$Nd/$^{238}$U and $^{232}$Th/$^{238}$U ratios in a specific phase. As we assume no fractionation occurs between Cm and Nd or Th, the initial ($t_0$) $^{247}$Cm/$^{144}$Nd and $^{247}$Cm/$^{232}$Th ratios are invariable between phases. Brennecka et al. (2010) and Tissot et al. (2016) have estimated the values for these ratios based on CAI data and we test both their values. From these ratios and an assumed formation age of 4567.3 Ma, the initial ($t_0$) $^{247}$Cm/$^{238}$U in the phase can be calculated through the following equations:

$$\left(\frac{^{247}Cm}{^{238}U}\right)_{phase,\,t_0} = \left(\frac{^{247}Cm}{^{144}Nd}\right)_{t_0} \times \left(\frac{^{144}Nd}{^{238}U}\right)_{phase,\,t} \times e^{-\lambda_{238}t} \qquad (1)$$

$$\left(\frac{^{247}Cm}{^{238}U}\right)_{phase,\,t_0} = \left(\frac{^{247}Cm}{^{232}Th}\right)_{t_0} \times \left(\frac{^{232}Th}{^{238}U}\right)_{phase,\,t} \times e^{(\lambda_{232}-\lambda_{238})t} \qquad (2)$$

A higher initial $^{247}$Cm/$^{238}$U in a given phase results in a higher present-day $^{235}$U/$^{238}$U ratio in that phase. In the Supplementary Materials, detailed descriptions can be found on the equations and parameter values used to calculate the resulting U isotopic composition. We report the results of the modelling in the $\varepsilon^{235}U$ notation, which is widely used to report U isotopic compositions. The value represents the parts per ten thousand deviations of the sample or mineral phase relative to the U isotopic composition of the standard CRM-145, which has a $^{238}U/^{235}U$ ratio of 137.844 (Condon et al., 2010):

$$\varepsilon^{235}U = \left( \frac{\left(\frac{^{235}U}{^{238}U}\right)_{phase,t}}{\left(\frac{^{235}U}{^{238}U}\right)_{CRM-145}} - 1 \right) \times 10000 \quad \text{or} \quad \varepsilon^{235}U = \left( \left( \left(\frac{^{235}U}{^{238}U}\right)_{phase,t} \times 137.844 \right) - 1 \right) \times 10000 \qquad (3)$$

# 3. Results

## *3.1 Mineralogy and petrography*

The inclusions have coarse-grained igneous textures and are mainly composed of primary melilite, anorthite, Al,Ti-pyroxene and spinel (Figs. 2, S1, Table 1). Based on their modal mineralogy, all CAIs except #44 are classified as Type B inclusions. Type B1 CAIs have a thick outer mantle of gehlenitic melilite whereas anorthite, Al,Ti-pyroxene and spinel are largely confined to the interior. In contrast, Type B2 CAIs lack such a mantle and display a more uniform mineral distribution (Bullock et al., 2013). In our sample set, CAI L1B-1 belongs to Type B1 while CAIs #30, #55 and L1B-2 are Type B2 inclusions (Fig. 2). The abundant forsterite grains in the core of CAI #54 characterises this inclusion as a forsterite-bearing Type B CAI (FoB; Fig. 2). The high modal abundance of anorthite and Al,Ti-pyroxene relative to melilite indicate that CAI #44 is a Type C inclusion (Fig. 2). Perovskite is observed only in CAI #30, where it appears to be a secondary phase formed by oxidation of primary Al,Ti-pyroxene (Fig. S1). Apart from this secondary perovskite, melilite, Al,Ti-pyroxene and spinel are unaltered in all inclusions. Minor secondary nepheline and sodalite occur as alteration of anorthite. Nepheline is present in all inclusions and sodalite occurs in all CAIs of NWA 18679, consistent with the elevated Cl concentration in this host chondrite. In particular, anorthite near the periphery of CAI #44 has been largely replaced by sodalite (Fig. 2). All CAIs are enclosed by WL rims and a forsterite-rich accretionary rim (FAR) further surrounds CAIs #54, #55 and L1B-1 (Figs. 2, S1).

| Host chondrite | CAI name | Size ($mm^2$) | Type | Group | Mineralogy | Th (ng/g) | U (ng/g) | Th/U |
|---|---|---|---|---|---|---|---|---|
| NWA 18679 | #30 | 6.9 × 3.1 | B2 | II | mel, px, an, sp, pv | 292 ± 5 | 323 ± 6 | 0.9 |
| NWA 18679 | #44 | 5.4 × 4.2 | C | II | px, an, mel, sp, sod | 545 ± 9 | 30.5 ± 0.5 | 17.9 |
| NWA 18679 | #54 | 6.7 × 5.1 | FoB | V | px, mel, sp, fo, an | 353 ± 4 | 74.9 ± 0.8 | 4.7 |
| NWA 18679 | #55 | 11.8 × 7.3 | B2 | I | mel, px, an, sp | 220 ± 3 | 118 ± 2 | 1.9 |
| Leoville | L1B-1 | 9.9 × 7.9 | B1 | V | mel, px, an, sp, met | 314 ± 5 | 57.1 ± 1.1 | 5.5 |
| Leoville | L1B-2 | 4.3 × 3.3 | B2 | III | mel, px, an, sp, met | 303 ± 6 | 40.4 ± 1.5 | 7.5 |

**Table 1**: Descriptions of samples in size, type, REE group, main mineralogy, bulk Th and U concentrations (2SE uncertainties) and Th/U ratios. Abbreviations of minerals are Al,Ti-pyroxene (px), melilite (mel), anorthite (an), spinel (sp), perovskite (pv), forsterite (fo), Fe,Ni metal grains (met), sodalite (sod).

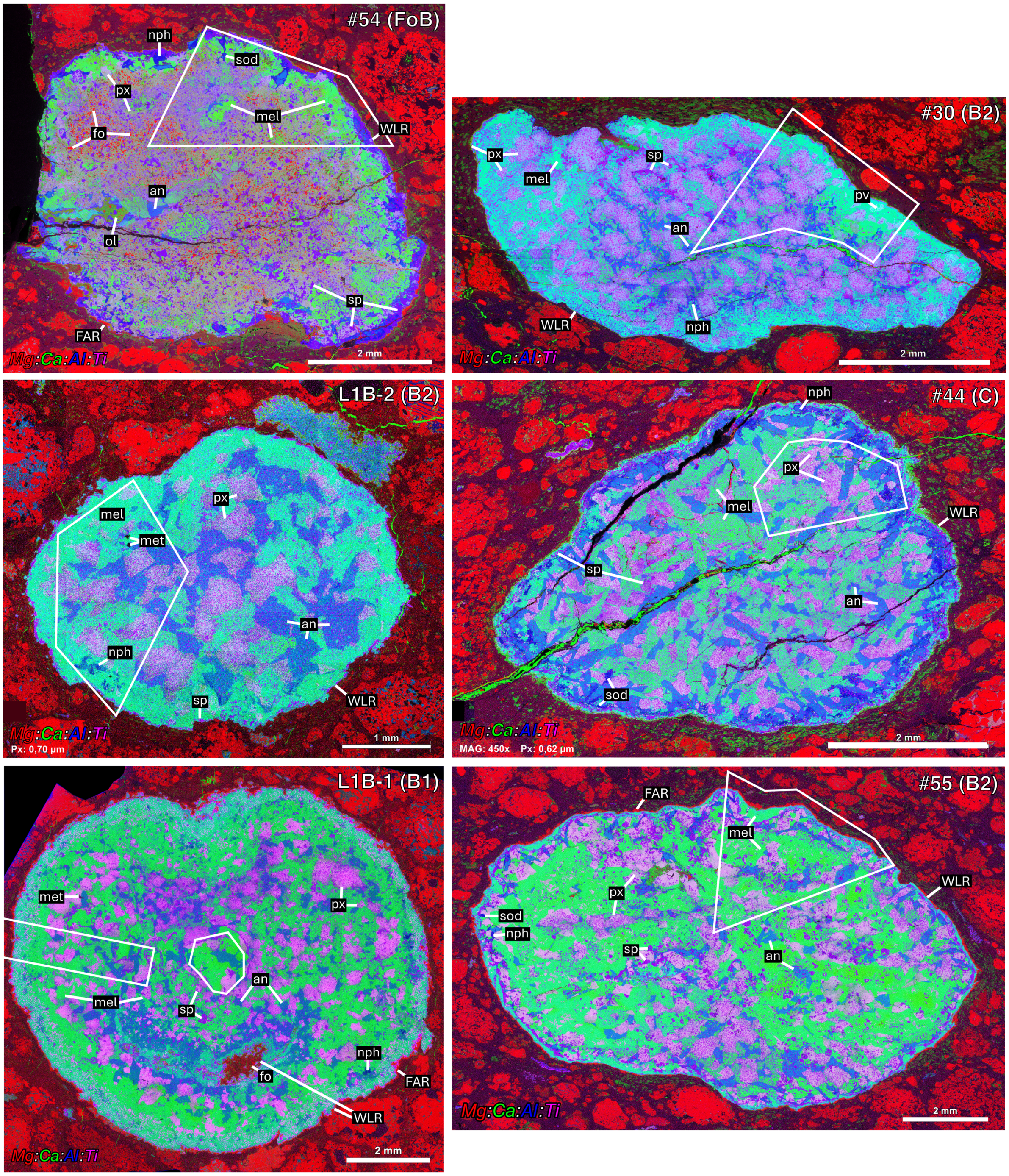

**Figure 2**: Combined EDS Mg:Ca:Al:Ti elemental maps of CAIs

Combined EDS elemental maps of Mg, Ca, Al and Ti in a red-green-blue-magenta colour scheme for the studied CAIs indicate major mineralogy and petrography. Abbreviations of mineral phases are Al,Ti-pyroxene (px), melilite (mel), anorthite (an), spinel (sp), perovskite (pv), forsterite (fo), Fe,Ni metal grains (met), olivine (ol), sodalite (sod), nepheline (nph). The inclusions are enclosed by Wark-Lovering rims (WLR), and some also exhibit an additional forsterite-rich accretionary rim (FAR). White polygons outline the regions targeted by fs-LA-ICP-MS mapping.

*3.2 Trace-element distributions and concentrations*

The regions of the CAIs targeted by fs-LA-ICP-MS are representative of their modal mineralogies and typically cover all textural zones (i.e. core, mantle, rim; Fig. 2). As such, we consider the average elemental concentrations from these regions to reflect their bulk chemistry (Table S1). The bulk REE patterns determined from these regions, reported relative to CI abundances, show distinct patterns that correspond to REE groups (Fig. 3; Mason & Martin, 1977). Inclusions #30 and #44 record a fractionated group II pattern, characterized by depletions in the most refractory heavy REEs (HREEs, except Tm and Yb) relative to the moderately refractory light REEs (LREEs). The other inclusions show flat, unfractionated REE patterns with consistent enrichment relative to the CI composition (10-18 × CI; Fig. 3). CAIs with small positive (#55) and negative (L1B-2) anomalies in the most volatile REEs (Eu, Yb and Ce) are classified, respectively, as group I and III. An absence of anomalies in these elements is characteristic of a group V pattern, recorded in #54 and L1B-1.

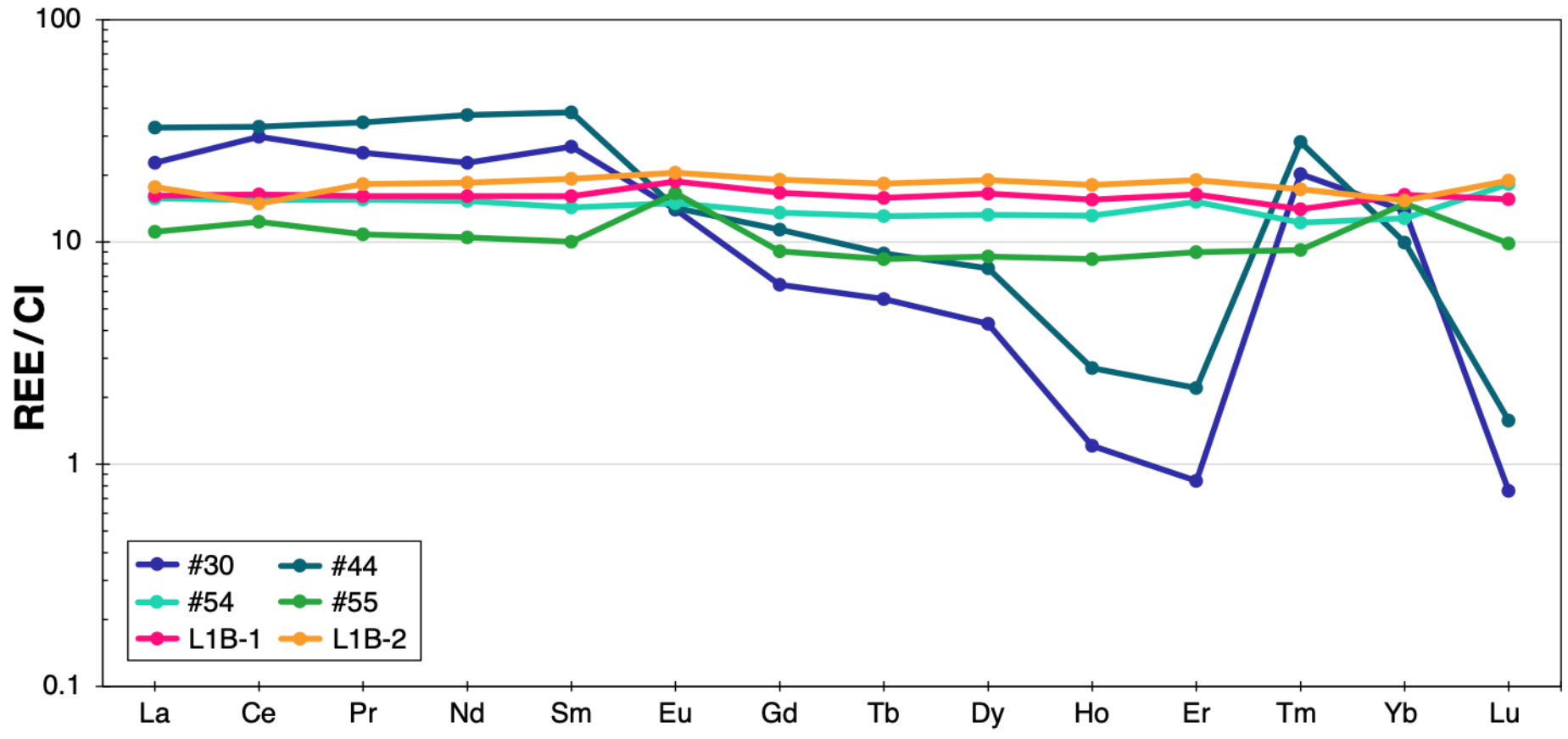

**Figure 3**: Bulk rare earth element patterns
Bulk rare earth element (REE) abundances, reported relative to CI chondrite abundances of Barrat et al. (2012), represent the average inclusions that were sampled by fs-LA-ICP-MS. Inclusions #30 and #44 show fractionated, group II REE patterns. The other inclusions show unfractionated REE patterns with consistent enrichment relative to CI composition. Slightly positive (#55) and negative (L1B-2) anomalies in the most volatile REEs (Ce, Eu, Yb) are characteristic of group I and III CAIs, respectively. Inclusions #54 and L1B-1 reveal no anomalies and are classified as group V CAIs.

Major- and trace-element distribution maps acquired by fs-LA-ICP-MS can be found in Supplementary Figures S2-8. The most relevant elemental maps for this study are presented in Figure 4 (i.e. Ti, Nd, Yb, Th, U). Trace elements are consistently hosted by the same mineral phases across the coarse-grained inclusions examined. Minor elements (V, Zr) and trace elements (Sc, Cr, U, Th, Pb, Y, REEs except Eu) are mainly hosted in Al,Ti-pyroxene. Where it can be distinguished, spinel is a significant host of V and Cr. Molybdenum is hosted primarily in metal, a feature that is particularly pronounced in the Leoville inclusions. The distributions of Sr and Eu are similar: both are mainly hosted in melilite and anorthite, although Sr is also enriched along fractures. Melilite is an important carrier of LREEs (La to Sm) and anorthite is the major Ba host with the highest concentrations near its grain rims (Figs. S6-8).

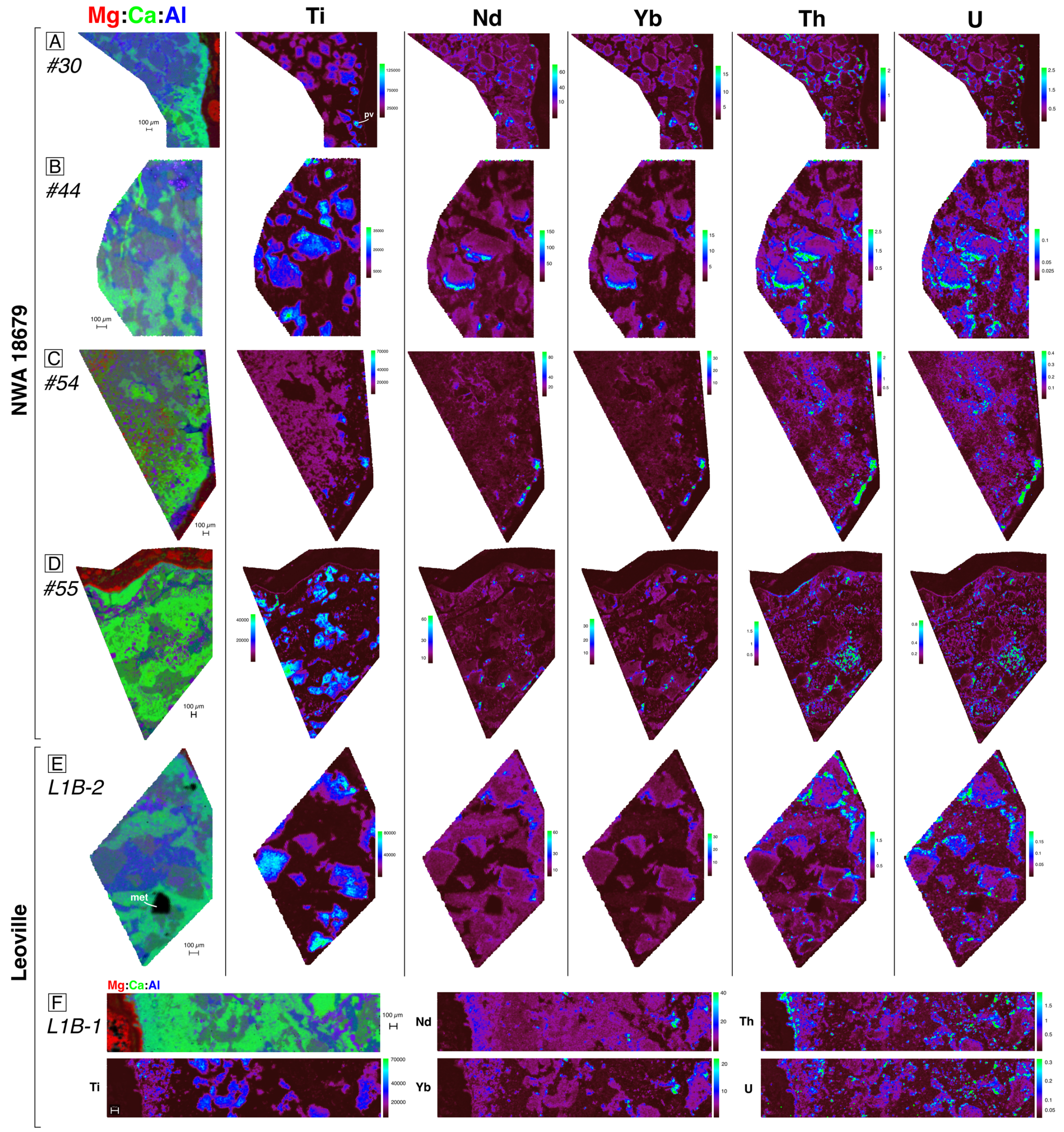

**Figure 4**: Elemental distribution maps by fs-LA-ICP-MS
Elemental maps acquired by fs-LA-ICP-MS of CAIs from NWA 18679 #30 (A); #44 (B); #54 (C); #55 (D) and Leoville L1B-2 (E); L1B-1 (F). All inclusions have a combined Mg-Ca-Al map from fs-LA-ICP-MS in RGB colour scheme, and elemental concentration maps in Ti, Nd, Yb, Th and U (in ppm).

Subdivision of the measured sections into distinct textural zones allows for the systematic evaluation of core-to-rim relationships. Particularly, refractory lithophile trace elements that are generally enriched in CAIs compared to the matrix, such as Y, REEs, Th and U, are enriched in the WL rims, and typically slightly enriched in the outermost layers relative to its bulk (Figs. 4; 5A). However, the most volatile REEs (Yb, Eu and Ce) show depletions in the WL rims and outer margins/mantle relative to more refractory REEs. This volatile REE depletion is expressed in Yb anomalies (Yb/Yb*) and appears coupled to Th/U ratios: from core to rim Yb anomalies decrease and Th/U ratios increase (Fig. 5B, C).

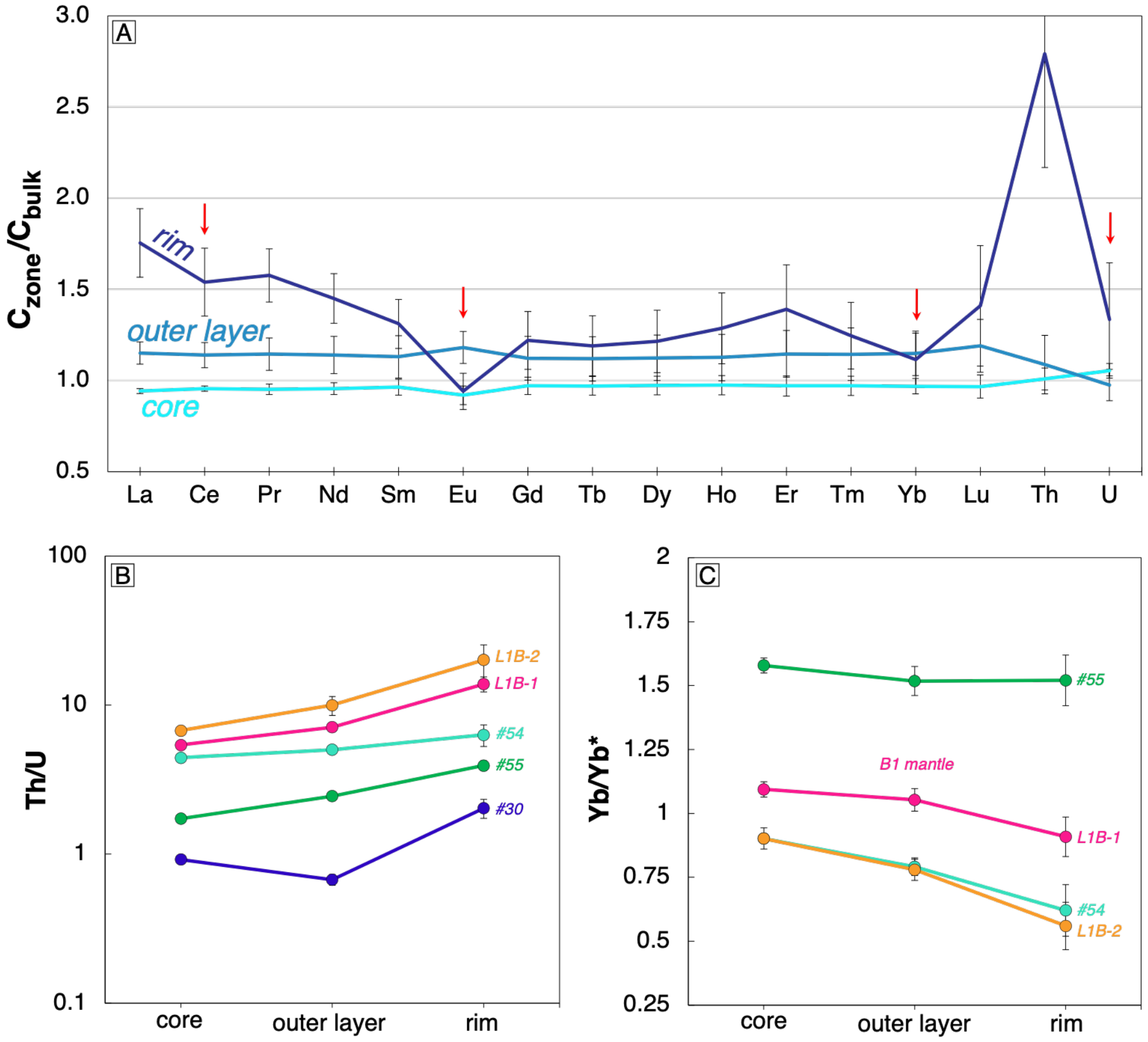

**Figure 5**: Core-to-rim trace element variations
(A) Bulk-normalized REE, U and Th abundances in each textural zone, averaged between CAIs #30, #54, #55, L1B-1, L1B-2, show general enrichment from core to rim. The rim patterns show relative enrichments in the most refractory REEs and Th, and depletions in the least refractory REEs (Ce, Eu, Yb) and U, indicated by red arrows. (B) Th/U ratios of individual inclusions increase from core to rim. (C) Yb anomalies ($Yb/Yb^* = Yb_N / (Tm_N \times Lu_N)^{0.5}$) of individual inclusions typically decrease from core to rim. Error bars represent 1SE (A) and 2SE (B, C).

### *3.3 Distribution of REEs within CAI minerals*

The measured REE patterns in CAI minerals reflect consistent partitioning behaviour amongst all inclusions, regardless of their bulk REE group (Fig. S9). All measured phases within individual inclusions exhibit similar group patterns that are modified in each phase through mineral-melt partitioning (Fig. 6). This mineral partitioning results in REE abundances of Al,Ti-pyroxene that increase with atomic number and REE abundances of melilite that decrease with atomic number. The REE patterns of melilite and Al,Ti-pyroxene typically cross around Ce, indicating melilite as main La host and Al,Ti-pyroxene as main host of REEs heavier than Ce, except Eu (Fig. 6). Strong positive Eu anomalies are recorded in anorthite and melilite and negative Eu anomalies in Al,Ti-pyroxene. Anorthite is a minor host for other REEs but shows a slightly stronger partitioning of LREEs than HREEs, like melilite. Analyses of the perovskite-rich region in CAI #30 demonstrate that perovskite is a major REE host, although perovskite is volumetrically minor in this inclusion. The degree of REE enrichment in perovskite cannot be resolved because the individual perovskite grains are smaller than the analytical spatial resolution. As a result, the reported concentrations reflect a mixture of perovskite and the surrounding melilite and Al,Ti-pyroxene. In contrast with Al,Ti-pyroxene, the REE pattern of the perovskite-rich region has a positive Yb anomaly but no Eu anomaly relative to the bulk inclusion.

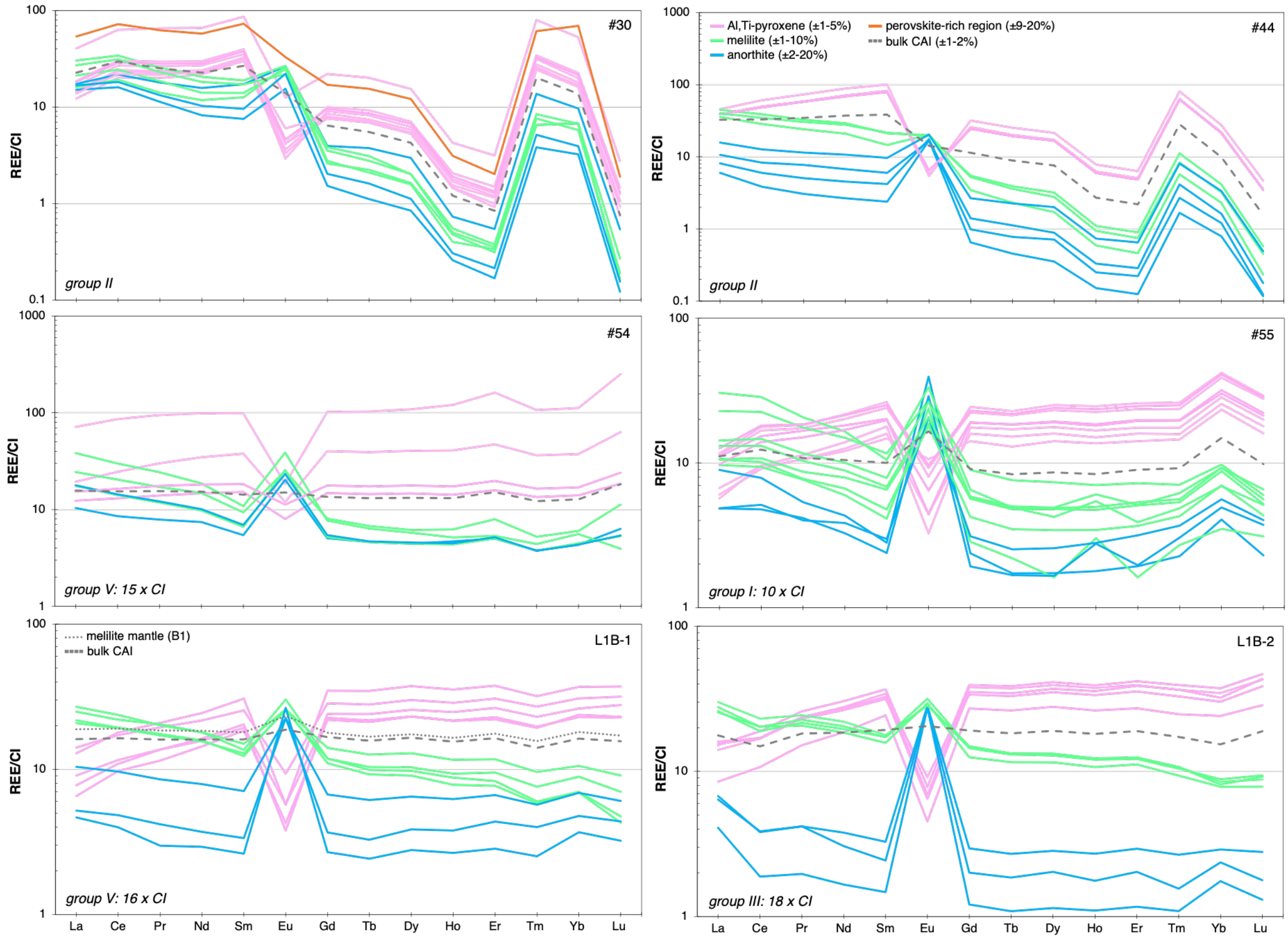

**Figure 6**: Phase-specific REE patterns in individual CAIs

The CI-normalized patterns show consistent REE partitioning between and within inclusions (CI chondrite abundances from Barrat et al. (2012)). Displayed data are obtained from selected regions of interest from fs-LA-ICP-MS maps for which averaged concentrations are reported in Table S1. Abundances in Al,Ti-pyroxene increase from LREE to HREE, with a pronounced negative Eu anomaly. The opposite pattern persists in melilite, which takes up higher abundances of LREE than HREE and has a positive Eu anomaly. Anorthite is a minor host of REEs, with decreasing abundances towards heavier REEs, although it has a strong positive Eu anomaly. In CAIs #30 and #44, the group II pattern presents in all phases, but similar mineral partitioning behaviour to non-group II CAIs is recorded for each phase, which supports an igneous origin of these inclusions.

The coarse-grained texture of the inclusions enables investigation of compositional distributions at the scale of individual mineral grains. Typically, the LREEs are homogeneously distributed throughout the melilite grains. In some grains, compositional variations in Al and Mg content (i.e. åkermanite number) indicate a higher abundance of LREE in gehlenitic melilite (Figs. S7-8). In Al,Ti-pyroxene, REE concentrations are inversely correlated with Ti, Zr, V and Sc content. Titanium concentrations decrease with increasing degrees of crystallization whilst REE abundances rise, resulting in the highest REE concentrations near the pyroxene grain edge (Figs. 4, 7).

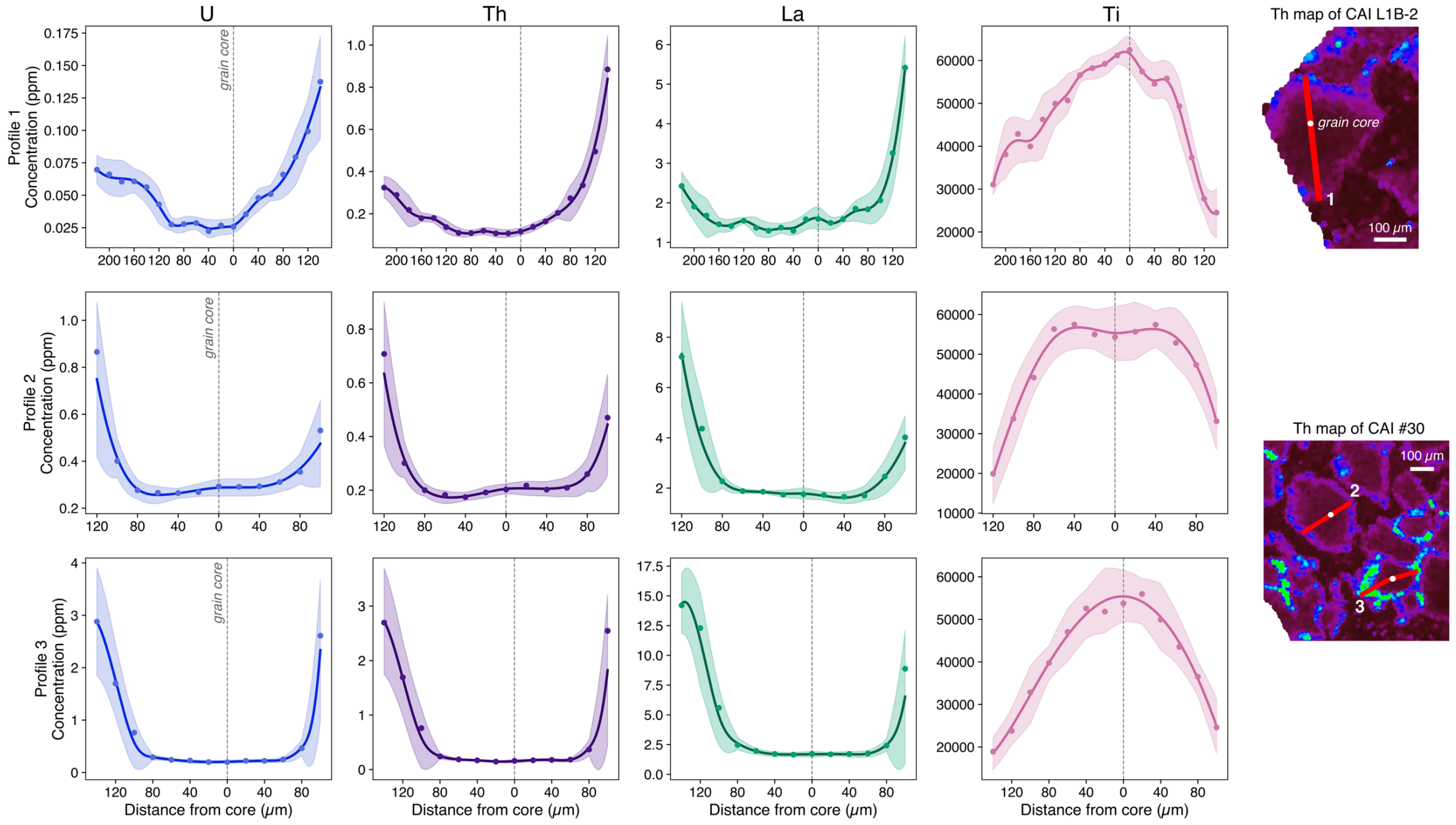

**Figure 7:** Cross-grain profiles of Al,Ti-pyroxene
Concentrations of U, Th, La and Ti (in ppm) are shown along three cross-grain profiles of Al,Ti-pyroxene. Individual points show the concentration in 20×20 $\mu m^2$ spots along the profile; the shaded background indicates the 2SE uncertainties on the points and the line shows the spline through the points. The Ti content is inversely correlated with U, Th and La abundances where Ti is most abundant in the grain core while U, Th and La rise in concentration toward the grain edge. Profiles are shown for Al,Ti-pyroxene grains in CAI L1B-2 (1) and CAI #30 (2, 3) on their Th distribution maps.

### *3.4 Distribution of U and Th within CAI minerals*

A wide range of U and Th concentrations is recorded in all major CAI minerals. This variability is coupled to the bulk concentrations within each CAI, that displays a unique Th-U relation (Fig. 8A). To identify the principal actinide host in each inclusion, Th and U abundances in CAI minerals are normalized to the bulk concentration of their host CAI (Fig. 8B).

This normalization approach reveals that Al,Ti-pyroxene typically exhibits the highest relative U and Th concentrations. The perovskite-rich region also records a higher U concentration than the bulk, but disproportionately records a lower Th concentration, as it plots below the 1:1 line that traces the bulk Th/U ratio of its host inclusion (Fig. 8B). This indicates an anomalously low Th/U ratio in this region, suggesting U partitions more efficiently into perovskite than Th. On the other hand, approximately one-third of melilite analyses plot above this 1:1 line, implying stronger partitioning of Th relative to U and consequently locally elevated Th/U ratios in melilite. Some spinel analyses also record elevated Th/U ratios, but their uncertainties are high given the small size of spinel-rich ROIs (Fig. 8B).

Within individual Al,Ti-pyroxene grains, distributions of U and Th mirror each other. Similar to the REE zoning within this mineral, the lowest U and Th concentrations are recorded where the Ti content is highest, whereas the highest U and Th concentrations are documented near or on the grain boundaries (Fig. 7). For example, the U concentrations in CAI #30 are 0.2-0.3 ppm in Al,Ti-pyroxene cores but rise to 1-3 ppm in the grain edge – a 10-fold increase (Fig. 7). The enrichment of U and Th in the Al,Ti-pyroxene grain edges is independent of the adjacent mineral phase. In some regions, distinguishing whether the actinides reside in the outermost grain edge of Al,Ti-pyroxene or in between grains is challenging. Where pyroxene grains are fractured (e.g. #44), U and Th concentrations are also elevated in the fractures (Fig. 4B). However, despite the fluid-mobile character of U, our samples show little evidence of U enrichment in larger Fe,Ni-rich cracks (Figs. S3, 4), attesting to the pristine nature of the inclusions in this study.

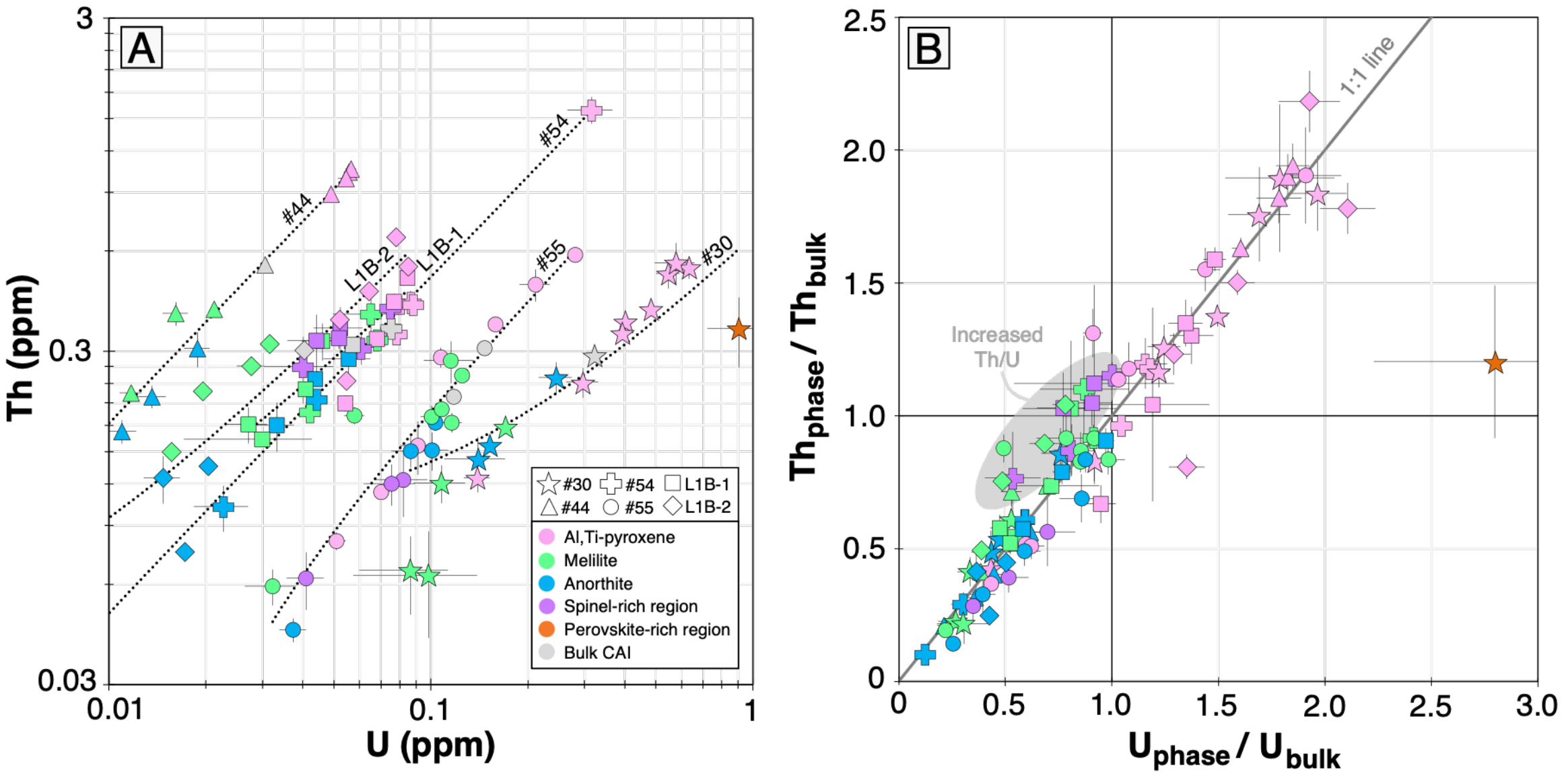

**Figure 8**: Th and U abundances in CAI minerals
(A) Compilation of Th and U concentrations (in ppm) in mineral phases of individual CAIs. The dotted lines are linear regression trends through the data points belonging to each individual CAI, as each inclusion has a specific Th-U concentration regime. (B) Th and U abundances in phases are normalized to their bulk concentrations to reveal Al,Ti-pyroxene as the main host phase for Th and U in all studied CAIs. Perovskite, which was only measured in one inclusion, is a significant host as well, particularly for U. Some analyses of melilite and spinel fall above the 1:1 line, indicating a stronger partitioning of Th than U into these phases, resulting an increased Th/U ratio. Error bars are 2SE.

### *3.5 Model inter-mineral U isotopic variability*

To show the feasible inter-mineral variation in U isotopic composition based solely on $^{247}$Cm decay, we show a model run using a Cm-free $^{238}U/^{235}U$ ratio of 137.844 (Fig. 9). This value represents $\varepsilon^{235}U = 0$ and allows us to exclusively show the effect of $^{235}$U ingrowth on the isotopic composition in $\varepsilon$-units (Fig. 9). The initial $^{247}Cm/^{238}U$ ratio in the phase, inferred from measured $^{144}Nd/^{238}U$ and $^{232}Th/^{238}U$ ratios, is allowed to evolve over 4567.3 Myr, thereby yielding a time-dependent adjustment to the Cm-free value. The $\varepsilon^{235}U$ model results are not sensitive to the exact evolution time within the age range proposed for CAI formation (conservatively 4567-4569 Myr), as virtually all $^{247}$Cm had decayed after 90 Myr.

Minerals with enrichment of Nd and Th relative to U carry the largest modelled $\varepsilon^{235}U$ excesses. Consistent with our previous results that show melilite is a major LREE host (incl. Nd; Fig. 6) and a phase shown to carry increased Th/U ratios (Fig. 8B), this mineral typically has the largest $\varepsilon^{235}U$ excesses in our model results (Fig. 9; Table S2). Anorthite records the largest $\varepsilon^{235}U$ deficits compared to the bulk inclusion, whereas Al,Ti-pyroxene has an isotopic composition close to the bulk (Fig. 9). The latter is expected because Al,Ti-pyroxene hosts the majority the U and Th budgets and is an important carrier of REEs and, therefore, this phase typically reflects the bulk composition. Of the two proxies for Cm, Nd predicts larger inter-mineral U isotopic variations than Th. Both, however, indicate CAIs #44 and L1B-2 as the inclusions with the largest variability in U isotopic compositions amongst their coexisting minerals. In the most extreme case of CAI #44, anorthite and melilite vary by ~7 $\varepsilon$-units using Nd as proxy while Th predicts only ~1 $\varepsilon$-unit of variation (Fig. 9).

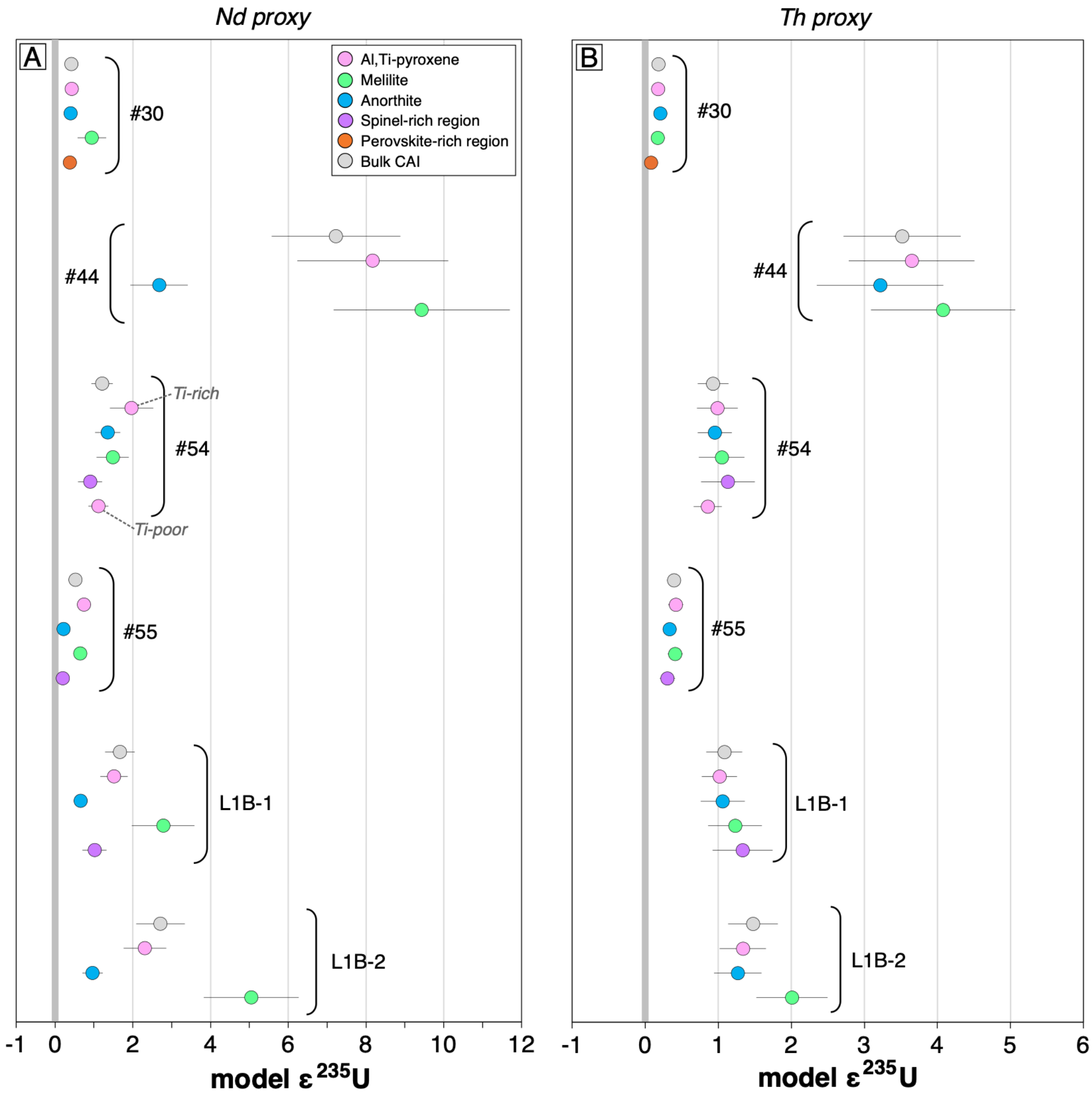

**Figure 9**: Modelled U isotopic variability between mineral phases of individual CAIs
Model results are reported in $\varepsilon^{235}$U values using Nd (A) and Th (B) as proxies for Cm. A Cm-free $^{238}$U/$^{235}$U ratio of 137.844 is used, to which a $^{235}$U excess is added through $^{247}$Cm decay. As this ratio is equal to the composition of the standard to which the results are normalized, the $\varepsilon^{235}$U deviation from 0 (grey line) solely reflects $^{247}$Cm decay. The $\varepsilon^{235}$U variability between phases is largest in CAIs #44 and L1B-2 and is most extensive using Nd as proxy for Cm. Error bars represent propagated uncertainties of our concentration data, initial $^{247}$Cm/$^{144}$Nd and $^{247}$Cm/$^{232}$Th ratios from Tissot et al. (2016) and uncertainties on decay constants (Table S5).

## 4. Discussion

### *4.1 Evaluation of fs-LA-ICP-MS mapping analysis*

The fs-LA-ICP-MS mapping approach employed in this study was designed to balance spatial resolution, analytical sensitivity, and areal coverage in order to characterise trace-element distributions at the scale of entire coarse-grained CAIs. Although the spatial resolution achieved here (20 µm spot size) is lower than that attainable by nanoSIMS or other ion-microprobe techniques, it is well matched to the coarse-grained textures of Type B and C CAIs and enables rapid, statistically robust mapping of large areas that capture representative mineralogical and textural variability. Higher resolution techniques are better suited for resolving sub-10 µm phases (e.g. spinel, perovskite) but this comes at the expense of areal coverage, highlighting the value of their complementary use alongside fs-LA-ICP-MS.

The present approach provides a substantial increase in analytical throughput compared to previous trace-element mapping studies of CAIs. Data acquisition rates averaged ~1.2 $mm^2$ $h^{-1}$, allowing mapped regions to cover approximately 6–18 % of the exposed CAI surfaces in polished sections. This sampling rate represents an order-of-magnitude improvement relative to prior fs-LA-ICP-MS mapping of fine-grained CAIs (Masuda et al., 2024) and several orders of magnitude improvement relative to nanoSIMS imaging of trace elements in coarse-grained inclusions (Ito & Messenger, 2016). Such coverage is critical for assessing bulk-representative behaviour and for evaluating the spatial context of trace-element zoning, particularly in minerals controlling the U and Th budgets.

### *4.2 Trace-element signatures of condensation, igneous partitioning and melt evaporation*

Our trace-element distribution maps archive the effects of multiple processes operating during the evolution of coarse-grained igneous CAIs, including condensation, evaporation and melting. The bulk trace element compositions were established by their initial formation through condensation and (partial) evaporation processes, which leave distinct imprints on REEs (Boynton, 1975; Mason & Martin, 1977; Hu et al., 2021). Group II REE patterns are recorded in two inclusions of this study and are thought to reflect volatility-controlled fractionation during condensation following evaporation (Hu et al., 2021). Fine-grained CAIs that have escaped melting virtually all have group II patterns and preserve their initial formation history, confirming their origin as nebular condensates (Boynton, 1975; Davis et al., 2018; Hu et al., 2021). In contrast, coarse-grained CAIs with a group II pattern are rare but typically spinel-rich (MacPherson et al., 2020), as is the case for our group II CAIs #30 and #44.

Mineral-specific partitioning of trace elements is consistent across all measured inclusions and is in line with previous *in situ* measurements of trace-element abundances in coarse-grained CAI minerals obtained by ion microprobe (e.g. Simon et al., 1991; 1999; Yi et al., 2024) and single-spot LA-ICP-MS (e.g. Campbell et al., 2003; Chaumard et al., 2014). We quantify the observed partitioning by normalizing mineral trace element abundances to the bulk abundance in the inclusion, essentially calculating mineral-melt partition coefficients (Table S3; Fig. S9). These coefficients quantify the distribution of trace elements between crystallizing phases and coexisting melts and have been experimentally determined for various minerals using CAI-like melts (Kuehner et al., 1989; Beckett et al., 1990; Simon et al., 1991; Davis et al., 1996; Lundstrom et al, 2006; Ustunisik et al., 2019). Our observed partitioning of trace elements into CAI minerals (e.g. LREE enrichment in melilite; Sc, Zr and Eu compatibility in Al,Ti-pyroxene; Fig. S9) agrees with the experimentally derived coefficients which indicates that the partitioning is consistent with igneous crystallization from a melt (Table S3). A deviation from previous studies concerns the behaviour of Th and U, which appear to partition similarly relative to one another across the investigated CAI minerals (Fig. S9).

Following melting, igneous CAIs largely behaved as closed systems, with efficient melt homogenisation and subsequent crystallisation following the expected igneous partitioning. For instance, all major minerals in group II CAIs #30 and #44 uniformly record the bulk group II signature (Fig. 6). In these inclusions, mineral compositions therefore reflect inheritance of the bulk signature that was modified by igneous processes (Fig. S9). Deviations from this behaviour require open-system processes such as high-temperature melt evaporation.

Clear evidence for melt evaporation is preserved in WL rims, which are highly enriched in REEs and Th (Wark & Boynton, 2001), although the most volatile REEs (Yb, Eu, Ce) and U are depleted relative to more refractory elements (Fig. 5A). This volatile REE and U depletion points to evaporative loss during the high-temperature event that formed the WL rims while their host CAIs were free-floating in the CAI-forming region, as WL rim formation occurred nearly contemporaneously with the formation of their host CAIs (MacPherson et al., 2026). Redox-dependent behaviour of volatile REEs and U may have contributed to their depletion in the WL rims, particularly under increasingly oxidizing conditions, which enhance the volatility of Ce and U (Davis et al., 1982; Wang et al., 2001; Wood et al., 2019). Both their redox sensitivity and lower 50% condensation temperatures ($T_{50\%} < 1550$ K; Wood et al., 2019) cause Eu, Yb and Ce to behave differently from most other REEs. Elevated Th/U ratios in the WL rims likely reflect the greater volatility of U relative to Th and are best explained by evaporative fractionation (Fig. 5B; Wood et al., 2019). Alternatively, the high Th/U in the rims could reflect that fluid-mobile U migrated from the CAI rim into the matrix post accretion, increasing local Th/U ratios. However, U concentrations in the matrix material near the rim are consistent with chondrite matrix values (~8-10 ppb U; Barrat et al., 2012).

The volatile-depleted, trace-element enriched WL rims are thought to be produced by intense (flash) heating of the outermost regions of their host CAI and/or recondensation onto the CAI (Wark & Boynton, 2001; MacPherson et al., 2026). Our results show that the WL rim of group II CAI #30 retains the characteristic fractionated group II signature of its interior (Table S1). Thus, the high concentration of actinides and REEs in its WL rim suggests their derivation from previously interior material that was not evaporated from the CAI surface during the flash heating episode (Wark & Boynton, 2001). The most refractory elements, enriched in the evaporation residue, likely recrystallized into rim layers consisting of spinel+hibonite+perovskite and gehlenitic melilite, whereas the outer pyroxene ± forsterite rim layers may have recorded subsequent recondensation instead (MacPherson et al., 2026). These outer rim layers may represent a back-reaction of the evaporated gas with the solid or molten inclusion surface. The resolution obtained in this study cannot spatially resolve the WL rim layer that hosts the Th and REE enrichment but the presence of the group II pattern in the rim of CAI #30 suggests this host layer was sourced from the melt residue.

The signatures of melt evaporation are also preserved in the outer regions of igneous CAIs, below their WL rim (Fig. 5; Wark & Boynton, 2001). In Type B1 inclusions, melilite mantles are depleted in relatively volatile elements such as Mg and Si, most likely due to melt evaporation (e.g. Davis et al., 1990; Bullock et al., 2013). The melilite mantle CAI L1B-1 of this study records decreasing Yb anomalies from core to rim (Fig. 5). In Type B2 CAIs, a melilite mantle is absent but we find that their outermost layers (200 µm subrim) also record a history of melt evaporation. The inclusions show increased Th/U ratios in this outer region compared to their cores and two inclusions record decreasing Yb anomalies from core to rim (Fig. 5B, C). Thus, their outer margins were likely modified by one or more thermal events that preferentially vaporized the most volatile REEs and U, relative to more refractory elements. These observations demonstrate that the thermal processing of Type B CAIs can generate evaporation-induced refractory enrichment at the melt margins, independent of whether they hold melilite mantles or not. Previous work has indicated that slower cooling rates and high hydrogen gas pressure ($P_{H2} > 1$ Pa; Kamibayashi et al., 2021) promote melilite mantle formation and enhance evaporation rates (Mendybaev et al., 2006). Variation in these conditions may control the degree of melt evaporation which could inhibit elemental diffusion within the melt and prevent full chemical homogenization (Richter et al., 2002; Mendybaev et al., 2006), producing a compositionally variable CAI population.

*4.3 Fractional crystallization causes mineral-scale zoning effects*

The extensive REE, U and Th zoning within Al,Ti-pyroxene grains (Fig. 7) can be explained through simple fractional crystallization, where the residual melt became increasingly enriched in incompatible trace elements such as REEs, U, Th as well as now-extinct Cm. According to the predicted crystallization sequence for Type B melts, pyroxene and anorthite crystallize after spinel and melilite (Stolper, 1982; MacPherson, 2014). When pyroxene starts forming, the melt is mainly enriched in HREEs, as LREEs have previously partitioned into melilite (Ito & Messenger, 2016), resulting in the observed pattern. This REE zoning pattern within pyroxene has also been measured by nanoSIMS and ion microprobe in Al,Ti-pyroxene of Type A and B CAIs (Simon et al., 1991; Kennedy et al., 1997; Ito & Messenger, 2016) and is also documented for actinides U and Th (Simon et al., 1991). As the highly enriched melt finally crystallizes, U and Th are incorporated into the last crystallization stage - the pyroxene grain edge (Fig. 7). Similarly, zoning effects of REEs and actinides possibly exist within anorthite, although the concentrations are likely too low for our method to resolve variations within anorthite grains.

Our distribution patterns suggest U and Th are more incompatible than the REEs in Al,Ti-pyroxene, as the measured shift in concentration from grain core to edge is more abrupt for U and Th than the REEs (Fig. 4). The mirrored distribution of U and Th within pyroxene grains suggests that these elements have similar pyroxene-melt partition coefficients, in contrast to findings by Simon et al. (1991) (Fig. S9). Higher spatial-resolution measurements would likely resolve even higher U and Th concentrations in the outermost grain edges than attainable with our fs-LA-ICP-MS method.

Although $^{247}Cm$ has fully decayed away, we can still estimate its initial distribution and the distribution of its decay products $^{235}U$, and eventually $^{207}Pb$, within these igneous inclusions based on the suggested proxies for Cm during igneous processes. Assuming Cm behaved similarly to Nd (i.e. LREE; Boynton, 1978; Tissot et al., 2016) during igneous crystallization, it would have partitioned efficiently into melilite along with Al,Ti-pyroxene, where it would trace the grain edge. If instead Cm behaved more like Th (Brennecka et al., 2010), it would be largely incorporated into the pyroxene grain edge and only a minor Cm fraction would reside in melilite. In either case, Cm was likely igneously zoned within Al,Ti-pyroxene grains, resembling the zoning patterns observed for REEs, U and Th (Fig. 4).

A grain boundary enrichment of U and Th had previously been measured by fission track radiography of Allende CAIs (Murrell & Burnett, 1987), although this study identified both melilite and Al,Ti-pyroxene as the main actinide hosts in Type B CAIs. We recognize a stronger partitioning of Th than U into melilite from our results, but consistently observe Al,Ti-pyroxene as the main actinide carrier (Fig. 8B). Murrell and Burnett (1987) interpreted the grain boundary enrichment as a potential feature of alteration. However, our inclusions show no signs of alteration on the Al,Ti-pyroxene grain edges (Fig. S1), yet they record a similar grain boundary enrichment of U and Th, thereby excluding alteration as its formation pathway. Instead, our results support primary crystallization as the mechanism that produced the enrichment in late-stage pyroxene, as first proposed by Simon et al. (1991).

The extensive zonation of REEs and actinides in Al,Ti-pyroxenes masks the true trace-element inventory (Fig. 7). Estimating trace-element budgets in bulk samples from discrete spot analyses of constituent minerals may be problematic without knowledge of the distribution within minerals. To explain the measured bulk concentration, accessory phases that are enriched in trace elements but have not been identified within the sample, are often invoked (e.g. U budgets in Huykens et al., 2025; Yi et al., 2024). Spot analyses typically target the centre of grains to avoid overlap with other phases but may therefore overlook critical zoning effects (Fig. 7). This bias leads to a systematic underestimation of trace-element contents in Al,Ti-pyroxene in CAIs, demonstrating the importance of LA-ICP-MS mapping rather than spot analyses.

*4.4 Actinide behaviour in coarse-grained CAIs*

Previous studies have mainly focused on the behaviour of actinides within CAIs at the bulk scale (Boynton, 1978; Brennecka et al., 2010; Tissot et al., 2016) whereas recent studies investigate U and Th abundances and U isotope variability at mineral scale (Yi et al., 2024; Li, 2024). In this section, we discuss in detail how our distribution maps improve understanding of actinide behaviour within and among igneous CAIs.

Heterogeneity in Th/U ratios was discovered both within individual CAIs (Connelly et al., 2012; Blichert-Toft et al., 2020) and across the CAI population (Brennecka et al., 2010; Tissot et al., 2016). Group II CAIs exhibit the largest variability in bulk $^{232}Th/^{238}U$ ratios, ranging from 5.4 to 70.3, suggested to reflect nebular fractionation (Brennecka et al., 2010; Amelin et al., 2010). Our study shows that igneous CAIs, whether part of group II or not, also record significant deviations from the solar Th/U ratio (~3.9; Blichert-Toft et al., 2010). Our inclusions show variable Th/U ratios between constituent minerals and core-to-rim zones within a given CAI (Figs. 5B, 7B), and among them (Table 1). The consistent core-to-rim increase in Th/U ratio within igneous CAIs may reflect the preferential volatilization of U over Th from the melt as U is more volatile than Th under solar nebula conditions (Fig. 5B), rather than parent body or terrestrial alteration (e.g. Münker et al., 2025). The strong gradients in Th/U from core to rim makes this ratio a new potential tracer for melt evaporation, complementing its use as an indicator of nebular fractionation.

Our results further indicate that the behaviour of U in coarse-grained CAIs is coupled to both Th and Yb, reflecting the combined influence of igneous partitioning and volatility-controlled processes. In terms of mineral-melt partitioning, U broadly follows Th, resulting in similar spatial distributions within CAI minerals (Fig. 4). However, unlike Th, U also exhibits systematics that correlate with the most volatile REEs, particularly Yb. This U-Yb coupling is evident at the bulk scale. For instance, the group III CAI L1B-2 displays a negative Yb anomaly, low U concentrations, and a supra-solar Th/U ratio, whereas the group I CAI #55 shows positive anomalies in the most volatile REEs (Yb, Eu, Ce), elevated U concentrations, and sub-solar Th/U (Table 1; Fig. 3). One of the group II inclusions (#30) similarly records relatively high Yb abundances compared to other group II CAIs (e.g. Brennecka et al., 2010; Davis et al., 2018) and hosts the highest U concentration and lowest Th/U ratio among the samples studied. These relationships support previous suggestions that U and Yb behave similarly during condensation and evaporation processes under solar nebula conditions, resulting in coupled bulk abundances (Tissot et al., 2016).

Notably, the similarity in behaviour between U and Yb is not readily explained by volatility alone. The 50 % condensation temperatures of U and Yb differ by ~80 K, whereas those of U and Th differ by only ~20 K (Wood et al., 2019). This discrepancy suggests that additional factors influence U behaviour during CAI formation. One likely contributor is variable valence state under differing oxygen fugacity conditions. While Th is exclusively tetravalent, U can exist in multiple oxidation states (Blichert-Toft et al., 2020), and its volatility is expected to increase under more oxidizing conditions (Davis et al., 1982; Petaev et al., 2013). Similarly, Yb, Eu and Ce are sensitive to oxygen fugacity, causing these elements to fractionate from most other REEs (Davis et al., 1982; Wang et al., 2001; Wood et al., 2019). These observations suggest that the apparent coupling between U and Yb reflects the combined effects of volatility and redox-dependent speciation rather than simple condensation behaviour. We emphasize that this interpretation represents a working hypothesis consistent with the observed trace-element systematics, rather than a uniquely constrained mechanism.

The distribution of Cm has important implications for the expected mineral-scale U isotope variations with CAIs (Fig. 9). If no remobilization occurred following igneous crystallization, variations in $^{235}U/^{238}U$ are expected where the initial U distribution does not mirror the Cm distribution, reflecting Cm-U fractionation due to their differences in partitioning behaviour.

Both Nd and Th proxies for Cm typically predict the largest $^{235}U/^{238}U$ excesses in melilite and deficits in anorthite, induced by variable partitioning effects (Fig. 9).

In addition to radiogenic effects from the decay of $^{247}Cm$, U isotope heterogeneity among bulk CAIs is also generated by other processes. This is evident from the U isotopic compositions of some CAIs that are outside the expected evolution of $^{235}U/^{238}U$ from $^{247}Cm$ decay (Connelly et al., 2012; Tissot et al., 2016). Additional mechanisms such as kinetic processes, nuclear field shift effects and redox-dependent processes can induce U isotopic fractionation (Dauphas & Schauble, 2016). In CAIs, however, the variations are likely controlled by mass-dependent fractionation during kinetic processes such as condensation and evaporation (Connelly et al., 2012; Tissot et al., 2016). Both condensation and evaporation preferentially concentrate the light $^{235}U$ isotope into the condensate or evaporated gas (Tissot et al., 2016). Consistent with this behaviour, coarse-grained CAIs typically record heavier bulk U isotope signatures than their fine-grained counterparts (Brennecka et al., 2010; Amelin et al., 2010; Connelly et al., 2012; Tissot et al., 2016), and several are heavier than the solar U isotope composition (Connelly et al., 2017). Preferential volatilization of light $^{235}U$ during evaporation events would have resulted in a heavy U isotope enrichment in the melt residue from which igneous CAIs crystallized (Tissot et al., 2016). Furthermore, the bulk U isotopic compositions of igneous CAIs correlate with Mg isotopes, following the theoretical Rayleigh fractionation line that is expected from isotopic fractionation during melt evaporation (Shollenberger et al., 2019).

Indeed, the only four CAIs with U-corrected Pb-Pb ages (22E, 31E, 32E, SJ-101) exhibit bulk U isotopic compositions that are closely linked to their petrography and formation histories. Fine-grained CAI 22E, interpreted as a gas-solid condensate, records $^{235}U$ enrichments associated with condensation, whereas coarse-grained CAI 32E displays an isotopically heavy U composition consistent with melt evaporation (Connelly et al., 2012). The heavy U isotope composition of SJ-101, a FoB CAI (Amelin et al., 2010), cannot be entirely attributed to evaporative loss as it experienced limited melting (Bullock et al., 2012), and may have inherited its heavy composition during condensation from a $^{235}U$-depleted gas. In contrast, the $^{235}U/^{238}U$ ratio of coarse-grained CAI 31E follows the expected evolution from $^{247}Cm$ decay and requires no substantial mass-dependent fractionation to explain its composition (Connelly et al., 2012).

In our proposed framework, kinetic mass-dependent fractionation primarily controls bulk U isotope compositions in igneous CAIs, whereas radiogenic ingrowth from $^{247}Cm$ decay may control mineral-scale U isotope variability instead. The interplay between these contributions provides critical context for interpreting U isotope systematics and their implications for U-Pb chronometry of CAIs.

### *4.5 Implications for high-precision U-Pb chronology of CAIs*

#### *4.5.1 Distribution of U*

Knowledge of the distribution of U and radiogenic Pb, and their concentrations and mineral carriers in CAIs can provide support in the interpretation of U-corrected Pb-Pb ages. To this end, Yi et al. (2024) measured U concentrations in the minerals of four coarse-grained CAIs by SHRIMP. Their results indicate the major minerals carry 10-20 times lower U concentrations than the measured bulk U content, leading them to suggest most U is hosted in unknown minor minerals (apart from perovskite). Instead, we propose that Al,Ti-pyroxene is the major U host, but its strong zoning effects hide the existing U inventory from biased methods that only target grain cores.

The absolute chronology of CAIs relies on the Pb-Pb dating technique that consists of progressive dissolution of bulk samples with various acids that increase in molarity (Connelly & Bizzarro, 2009). The most radiogenic (i.e. highest $^{206}Pb/^{204}Pb$) fraction is typically collected at the first exposure of the CAI to hydrofluoric acid (Connelly et al., 2012), an acid that preferentially attacks silicates and oxides. The release of radiogenic Pb by the first hydrofluoric acid attack agrees well with the observed distribution of U and, consequently, its decay products $^{206}Pb$ and $^{207}Pb$ near Al,Ti-pyroxene grain edges. One possible explanation is that this first leach preferentially accesses the U-rich regions near exposed pyroxene surfaces. However, the grain-size distribution and fracture behaviour of crushed CAI material are not well constrained, making it difficult to determine which mineral surfaces are exposed during the dissolution procedure. Nevertheless, if U were predominantly concentrated in pyroxene grain cores rather than grain edges, one might expect the most radiogenic Pb to be released in subsequent dissolution steps or the residue because the material disintegrates with increased acid exposure.

Further, the grain edge enrichment of U in Al,Ti-pyroxene also makes this U more susceptible to mobilization during aqueous alteration, especially in regions where circulating fluids intrude along cracks. Although pyroxene is among the most chemically robust CAI minerals (Krot et al., 2021), alteration could redistribute fluid-mobile U and Pb away from their original sites. Therefore, U-Pb formation ages of coarse-grained CAIs are best determined from samples like ours that have experienced minimal aqueous alteration, ensuring that Pb-Pb ages reflect their crystallization rather than later disturbance.

*4.5.2 Inter-mineral U isotope variation*

Our model predicts that melilite records the largest $\varepsilon^{235}U$ excesses, owing to its stronger affinity for Cm than U (Fig. 9). In support of this model result, Li (2024) reported a resolved $^{235}U/^{238}U$ of melilite from its bulk host inclusion (0.47 ± 0.21 ‰), which is a larger difference than previously found (Shollenberger et al., 2019). This CAI is an exceptionally large, Type B inclusion 'AI01' from CV chondrite NWA 6870 – similar to the CAIs measured in this study. Thus, we propose that the resolved $^{235}U/^{238}U$ excess in melilite by Li (2024) is the result of variable Cm-U partitioning among coexisting minerals. Other mechanisms to explain their data could involve (1) U isotope heterogeneity in the precursor materials or, (2) U isotope fractionation during melt crystallization. We consider both mechanisms unlikely because (1) igneous CAIs experienced nearly complete melting that homogenized the composition prior to crystallization and, (2) heavy U isotopes are difficult to fractionate between minerals during crystallization at the rapid cooling rates of igneous CAIs (0.5 – 50 K/hr; Stolper & Paque, 1986; Yamamoto et al., 2021). Yet, U isotope fractionation during crystallization was recently suggested to explain inter-mineral U isotope variability recorded in plutonic angrites that form at much slower cooling rates than CAIs and volcanic angrites (Tissot et al., 2017; Huyskens et al., 2025). Contrary to plutonic angrites, volcanic ones have homogeneous U isotope compositions among their constituent minerals, implying a control of cooling rate on the fractionation of U isotopes (Huyskens et al., 2025).

If a bulk sample was used for progressive dissolution, CAIs with extensive U isotope variations between coexisting minerals would record non-linearity in Pb isotope space. The Mean Square of Weighted Deviates (MSWD) of an isochron quantifies the scatter of points around the isochron and is commonly used as a measure of isochron quality. If U isotope variations within a sample cause non-linearity in Pb isotope space, the MSWD would increase beyond acceptable values. Good isochron linearity may therefore indicate limited U isotope variations within a CAI. However, in reality, this assessment is more complex because progressive dissolution does not perfectly separate individual minerals and each leachate may contain variable contributions from multiple minerals. To assess inter-mineral U isotope heterogeneity in CAIs using Pb isotope data, future work should quantify mineral contributions in leachate fractions or measure Pb isotopic compositions of pure mineral separates (e.g. Bouvier & Wadhwa, 2010).

Despite this limitation, points plotting above the inverse Pb-Pb isochron could be explained by inter-mineral U isotope heterogeneity. For CAIs, these points above could be caused by the dissolution of melilite, which records $\varepsilon^{235}$U excesses that result in higher radiogenic $^{207}$Pb$^{*}$/$^{206}$Pb$^{*}$ signatures. Leachates plotting above the isochron have previously been attributed to leaching-induced fractionation of $^{207}$Pb$^{*}$ and $^{206}$Pb$^{*}$ (Amelin et al., 2025) but could also be related to U isotope variations between constituent minerals. Further, the production of $^{208}$Pb through $^{232}$Th decay makes $^{208}$Pb/$^{206}$Pb ratios of leachates valuable for constraining the Th/U variations within a CAI (Blichert-Toft et al., 2020). Therefore, $^{208}$Pb/$^{206}$Pb ratios may correlate with Cm abundance and U isotopic composition, providing another useful proxy for assessing U isotope variations within CAIs.

*4.5.3 Age offsets*

Inter-mineral heterogeneity in U isotopic composition leads to apparent age offsets in U-corrected Pb-Pb chronometry when a uniform U isotope ratio is assumed. To evaluate the magnitude of such offsets in the studied CAIs, we use the modelled mineral-specific U isotopic compositions (Fig. 9) to calculate the corresponding radiogenic Pb components. For each mineral phase, the present-day radiogenic $^{207}$Pb$^{*}$/$^{206}$Pb$^{*}$ ratio, assuming CAI formation at 4567.3 Ma (Connelly et al., 2012) is given by:

$$\left(\frac{^{207}Pb^{*}}{^{206}Pb^{*}}\right)_{phase} = \left(\frac{^{235}U}{^{238}U}\right)_{phase} \times \frac{e^{\lambda_{235}t}-1}{e^{\lambda_{238}t}-1} \qquad (4)$$

To assess possible age offsets, we explore scenarios in which the radiogenic Pb derives from one mineral phase while the U isotopic composition is taken from another (Fig. 10). Such mismatching of isotopically distinct reservoirs produces offsets from the 'real' age (i.e. 4567.3 Ma; $t_1$) used for the calculation to derive the U and Pb source phase compositions. For a given mismatched pair – e.g. radiogenic Pb from phase A and U from phase B – the resulting age $t_2$ is obtained by solving:

$$\left(\frac{^{207}Pb^*}{^{206}Pb^*}\right)_{phase\ A,t_1} = \left(\frac{^{235}U}{^{238}U}\right)_{phase\ B,t_1} \times \frac{e^{\lambda_{235}t_2}-1}{e^{\lambda_{238}t_2}-1} \qquad (5)$$

The resulting age offset $\Delta t$ is then:

$$\Delta t = t_1 - t_2 \qquad (6)$$

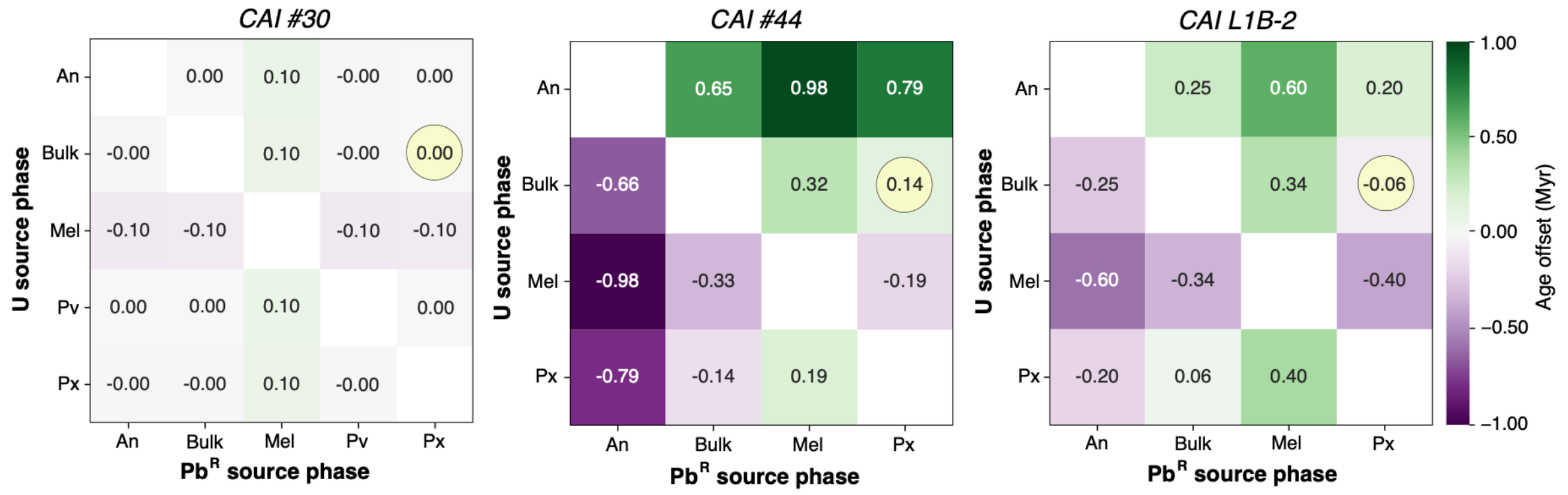

**Figure 10**: Calculated age offsets for CAIs #30, #44 and L1B-2 using Nd as proxy for Cm
The isotopic compositions of $Pb^R$ and U are calculated within each phase assuming a formation age of 4567.3 Myr (Connelly et al., 2012). By mismatching the source phases of U and radiogenic Pb and recalculating a hypothetical formation time, the inferred ages inform about potential age offsets resulting from the current methodology of U-corrected Pb-Pb dating. Yellow circles indicate the most realistic scenario, where the radiogenic Pb composition corresponds to pyroxene and the U composition to the bulk CAI.

These predicted age offsets are evaluated in the context of how U-corrected Pb-Pb ages of CAIs are typically obtained. In most studies, U isotopes are measured on bulk material, meaning the composition reflects a mixture of all constituent minerals. Lead isotopes are obtained from leachate fractions that correspond to points in Pb isotope space, through leaching of bulk samples (Amelin et al., 2010; Connelly et al., 2012; Li, 2024) or mineral separates (Connelly et al., 2008; Bouvier & Wadhwa, 2010). If standard isochron criteria are satisfied (Connelly et al., 2017), these points define an isochron that identifies the pure radiogenic Pb component which is used in the age calculations. However, the determination of this component is skewed towards the most radiogenic points (i.e. close to the y-axis intercept of an inverse Pb-Pb diagram) that typically reflect the most U-rich phase dissolved during bulk sample leaching. In the CAIs studied here, we identify Al,Ti-pyroxene as the major host of U (Fig. 8B), implying that the radiogenic Pb component represents pyroxene. This realistic mismatch scenario, where the bulk U isotopic composition is paired with an Al,Ti-pyroxene-dominated radiogenic Pb composition, introduces the possibility of erroneous ages if minerals differ significantly in their U isotopic composition.

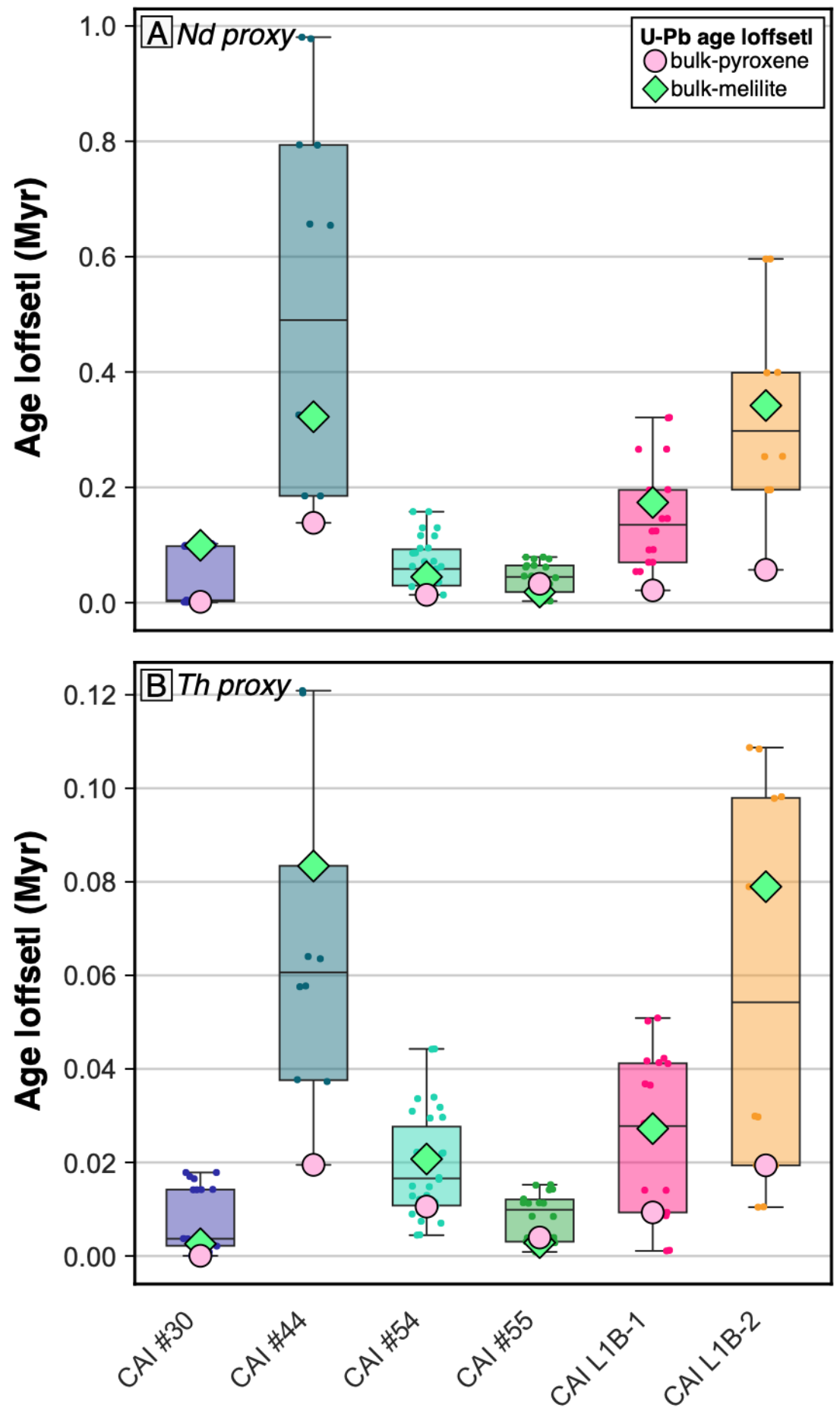

**Figure 11**: Absolute age offsets for all scenarios using Nd and Th as proxies for Cm
Boxplots of the age offsets (in absolute values) through cross-phase matching of U and $Pb^{R}$ sources, for each CAI using Nd (A) and Th (B) as proxies for Cm. The boxplot indicates the range in |age offsets| obtained from the mismatching scenarios (Fig. 10). The age offsets predicted by Th are ~5 to 10 times lower than the offsets predicted by Nd. For both proxies, the largest offsets are predicted for CAI #44 and L1B-2. Pink circles indicate the most realistic scenario in which the U and $Pb^{R}$ isotopic compositions derive from the bulk and pyroxene, respectively. Green diamonds indicate the offset in the scenario where U and $Pb^{R}$ isotopic compositions represent the bulk and melilite, respectively.

In general, Nd is regarded as the most reliable proxy for Cm (Tissot et al., 2016) and is particularly indicative during igneous processes (Boynton, 1978). As such, we discuss in detail the potential age offsets arising from inter-mineral U isotopic variations inferred from the Nd proxy, noting that the offsets from the Th proxy are smaller (Fig. 11; Table S4). As argued above, the scenario most consistent with current analytical practice assumes that the radiogenic Pb isotopic composition reflects pyroxene and the measured U isotopic composition represents the bulk inclusion. Under this scenario, the age offsets are minimal for most CAIs (0.00-0.03 Myr; Fig. 11), implying that phase-specific effects of $^{247}$Cm decay do not invalidate previously published U-corrected Pb-Pb ages of CAIs. However, caution should be taken in sampling the inclusions for U isotopic analyses. Where sufficient material exists, pyroxene separates should be used for both U and Pb isotope analyses. Alternatively, bulk $^{144}$Nd/$^{238}$U ratios can provide a first-order assessment of inter-mineral variability: inclusions with higher bulk ratios suggest greater potential influence of $^{247}$Cm decay on phase-specific U isotopic compositions. Taken together, these findings suggest that CAIs with limited variation in $^{144}$Nd/$^{238}$U among constituent phases are best suited for bulk-sample U-Pb chronometry.

In inclusions with large internal variations in $^{144}$Nd/$^{238}$U, however, the predicted age offsets become more significant. This large inter-mineral variability can be promoted by a high bulk $^{144}$Nd/$^{238}$U ratio (typically characteristic of volatility-fractionated, group II CAIs, e.g. CAI #44), or strong igneous partitioning effects (e.g. CAI L1B-2, Fig. 6). For example, CAIs L1B-2 and #44 show predicted age offsets of 0.06 and 0.14 Myr, respectively, even under the most realistic model scenario (Fig. 11A). These offsets approach the analytical uncertainty (±0.16 Myr) on the canonical absolute CAI formation age (Connelly et al., 2012), highlighting that variable Cm-U partitioning can meaningfully influence age interpretations and must therefore be accounted for in the associated uncertainty budget. If sufficient material is available, pyroxene can be separated from CAIs exhibiting such internal variations and analysed independently, thereby minimizing the effects of inter-mineral U isotope variations on the resulting age determinations.

The largest age offsets arise in scenarios involving melilite because this mineral hosts the greatest $\varepsilon^{235}$U excesses (Fig. 9). Melilite is both abundant in CAIs and characterized by moderate U concentrations, making it a key phase for assessing Cm-derived U isotope effects. In particular, the largest offsets, reaching up to ~1.0 Myr, are predicted from pairing of anorthite and melilite as the U and radiogenic Pb reservoirs, although this scenario is unrealistic (Fig. 10). Even so, these results underscore that CAIs with substantial melilite-pyroxene or melilite-anorthite partitioning differences require additional caution when interpreting U-Pb ages.

Finally, we calculate the potential age offsets resulting from the measured inter-mineral U isotope variations in CAI 'AI01', as these data represent direct observations rather than model outputs (Li, 2024). The resolved variation in $^{235}$U/$^{238}$U between its melilite and bulk inclusion is 4.7 ± 2.1 $\varepsilon$-units, translating to the same $\varepsilon$-variation for $^{207}$Pb$^{*}$/$^{206}$Pb$^{*}$. Scenarios in which the U isotope composition is represented by the bulk inclusion and the radiogenic Pb composition is represented by either melilite or Al,Ti-pyroxene show 0.68- and 0.22-Myr age offsets, respectively. These significant age offsets predicted for CAI 'AI01' further exemplify the necessity to derive U-corrected Pb-Pb ages from CAI mineral separates where possible. Analyses of the $^{144}$Nd/$^{238}$U and $^{232}$Th/$^{238}$U ratios in the mineral separates of this inclusion could reveal if the inter-mineral U isotope heterogeneity is related to variable Cm-U partitioning.

## 5. Conclusions

We have analysed trace-element abundances and distributions in six coarse-grained CAIs from $CV3_{red}$ chondrites using femtosecond laser ablation ICP-MS. This approach provides rapid, high-throughput measurements and statistically robust mapping over areas large enough to characterise entire inclusions, which is essential for evaluating bulk-representative behaviour and the spatial context of trace-element zoning.

Our results support a multi-stage history for igneous CAIs in which initial condensation-evaporation set their bulk trace element chemistry, and subsequent melting-crystallization episodes produced the mineral-scale trace-element distribution through igneous partitioning. During the melt stage, high-temperature evaporation caused volatile trace-element loss from the outer margins of CAIs, generating systematic shifts in Yb anomalies and Th/U ratios from core to rim. These gradients can serve as useful tracers of the degree of melt evaporation.

The dominant actinide (Th, U) and REE mineral carrier in coarse-grained CAIs is Al,Ti-pyroxene, if perovskite and hibonite are absent. Igneous zonation of REEs, Th, U, and by inference now-extinct Cm, in pyroxene grains can be explained by fractional crystallization from a residual melt enriched in incompatible elements. These zoning patterns are often overlooked due to the targeting bias of discrete spot analyses, emphasizing the value of laser-mapping approaches to capture complete trace-element inventories.

Variable Cm-U partitioning between CAI minerals produces inter-mineral U isotope variations, as predicted by $^{247}$Cm-$^{235}$U decay modelling, and can generate apparent age offsets in U-Pb chronometry when a uniform bulk U isotopic composition is assumed. Due to a stronger partitioning of Cm than U into melilite, this mineral hosts the largest $^{235}U/^{238}U$ excesses and can cause age offsets of up to ~0.30 Myr. Considering a scenario that aligns with the U-corrected Pb-Pb dating technique of CAIs, age offsets range from 0 to 0.14 Myr. The latter is comparable to the ± 0.16 Myr analytical uncertainty on the canonical absolute CAI formation age, implying some CAIs are inherently better suited for bulk-sample U-Pb dating than others. Alternatively, in CAIs characterized by substantial inter-mineral variability in U isotopic composition, pyroxene separates may be preferentially targeted for U-Pb dating to minimize the effects of U isotopic heterogeneity and obtain more reliable age determinations. These findings add further complexity to the determination and interpretation of absolute ages of the oldest Solar System solids.

**Acknowledgements**
We thank Audrey Bouvier for efficient handling of our manuscript and Andrew Davis and two anonymous reviewers for their insightful and constructive comments which improved this manuscript. We also thank Yuki Masuda, Jean Bollard and Jim Connelly for helpful discussions. Financial support for this project was provided by grants from the Villum Foundation (grant #54476 PLANETS) to M.B. and (Young Investigator grant #53024) to E.v.K.

**Declaration of generative AI and AI-assisted technologies in the manuscript preparation process**
During the preparation of this work the authors used Microsoft Copilot in order to improve the readability of the manuscript. After using this tool, the authors reviewed and edited the content as needed and take full responsibility for the content of the published article.

**Data availability**
Data are available through Mendeley Data at https://doi.org/10.17632/28thhdtkcz.1.

**Appendix A. Supplementary Material**
The supplementary files contain Figures S1-S9, detailed methods on Cm-U decay modelling and Tables S1-S5. The figures contain BSE images and all elemental maps obtained by fs-LA-ICP-MS of the CAIs. Tables include major and trace element abundances of bulk CAIs or individual mineral phases extracted from laser maps, results of Cm-U decay modelling and obtained age offsets.

Supplementary Materials

# Trace element distribution in coarse-grained Ca-Al-rich inclusions by femtosecond LA-ICP-MS: Implications for U-Pb chronology

M. E. Onderwater, E. M. M. E. van Kooten, G. J. MacPherson, A. N. Krot, M. Bizzarro

This file contains:
Supplementary Figures S1-S9
Supplementary Methods
Supplementary Table S5

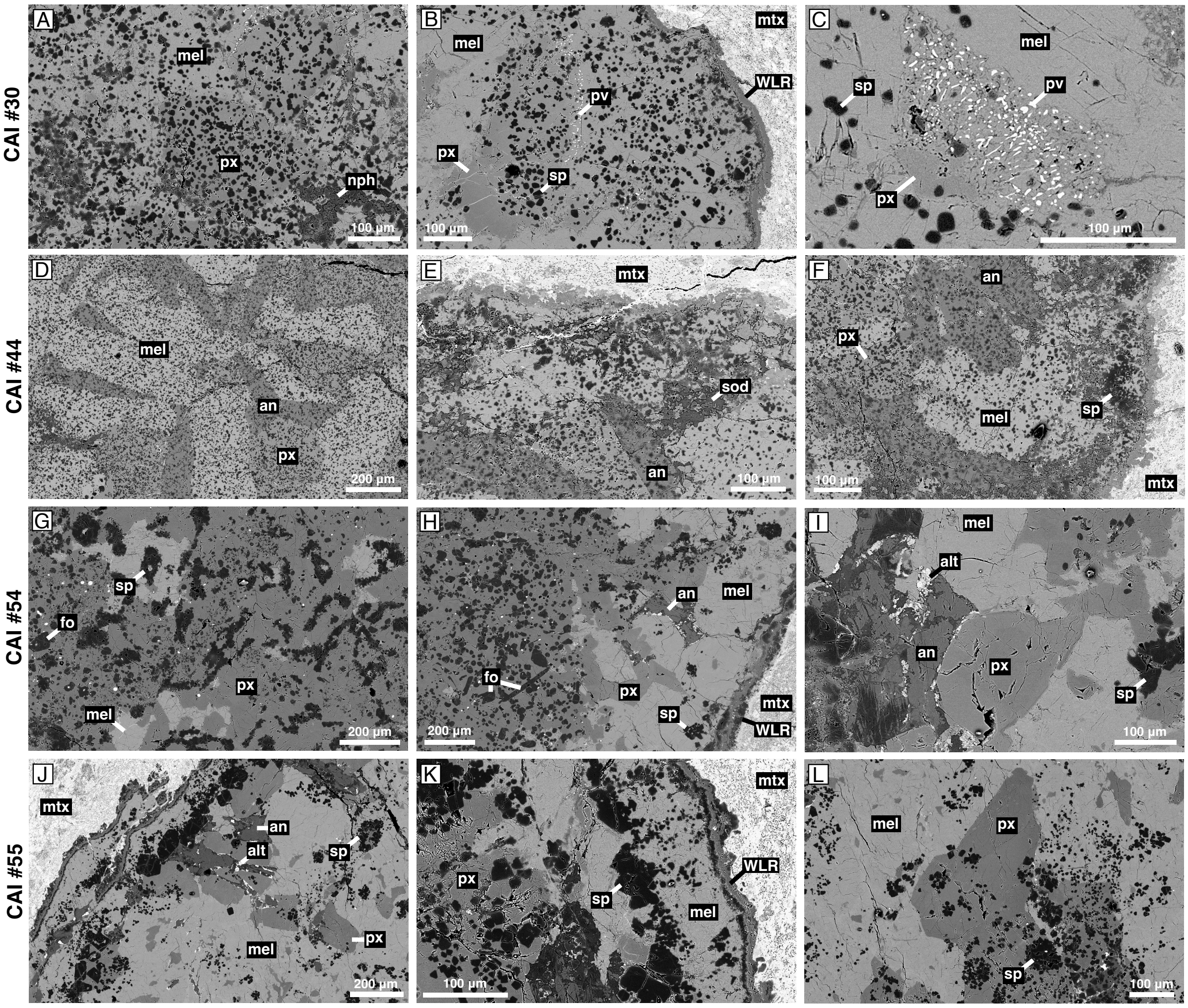

**Figure S1:** BSE images, acquired at the Smithsonian Institution, of NWA 18679 CAIs #30 (A-C), #44 (D-F), #54 (G-I), #55 (J-L). Abbreviations of mineral phases and features are pyroxene (px), melilite (mel), anorthite (an), spinel (sp), perovskite (pv), forsterite (fo), nepheline (nph), sodalite (sod), alteration (alt), Wark-Lovering rim (WLR) and chondrite matrix (mtx). Anorthite has experienced some alteration, mainly focused on its grain edges. Pyroxene grains are generally pristine, except in the outer region of CAI #30 where it has been oxidized to secondary perovskite (B, C). Abundant fine-grained spinel occurs in CAIs #30 and #44, while spinel nodules, palisades and coarser spinel grains are present in CAIs #54 and #55.

**Figure S2-8:** Major- and trace-element distribution maps of coarse-grained CAIs in this study, acquired by fs-LA-ICP-MS and reported in ppm (μg/g). The criteria maps show the distribution of mineral phases identified using Iolite classification criteria, with each colour representing a different mineral.

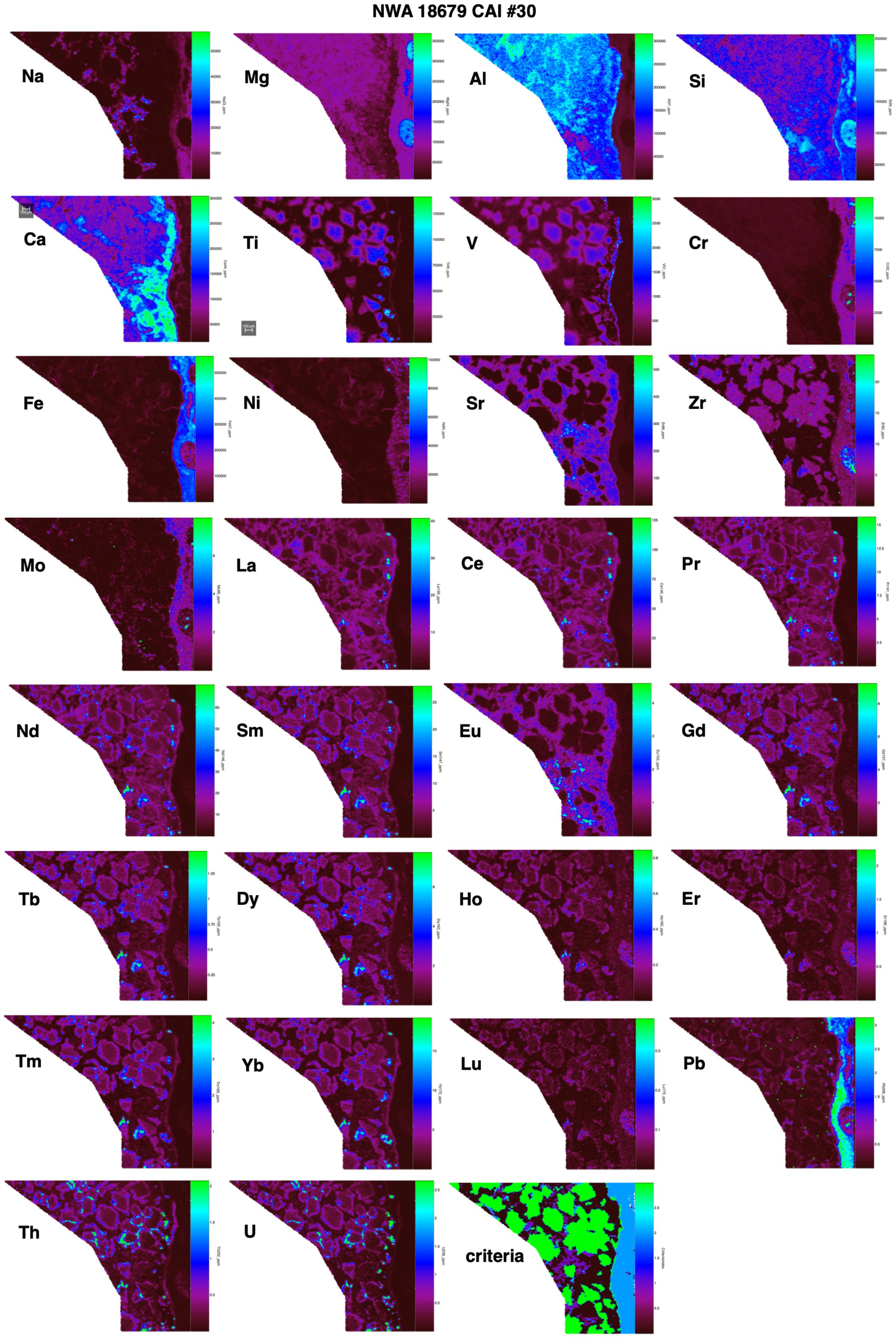

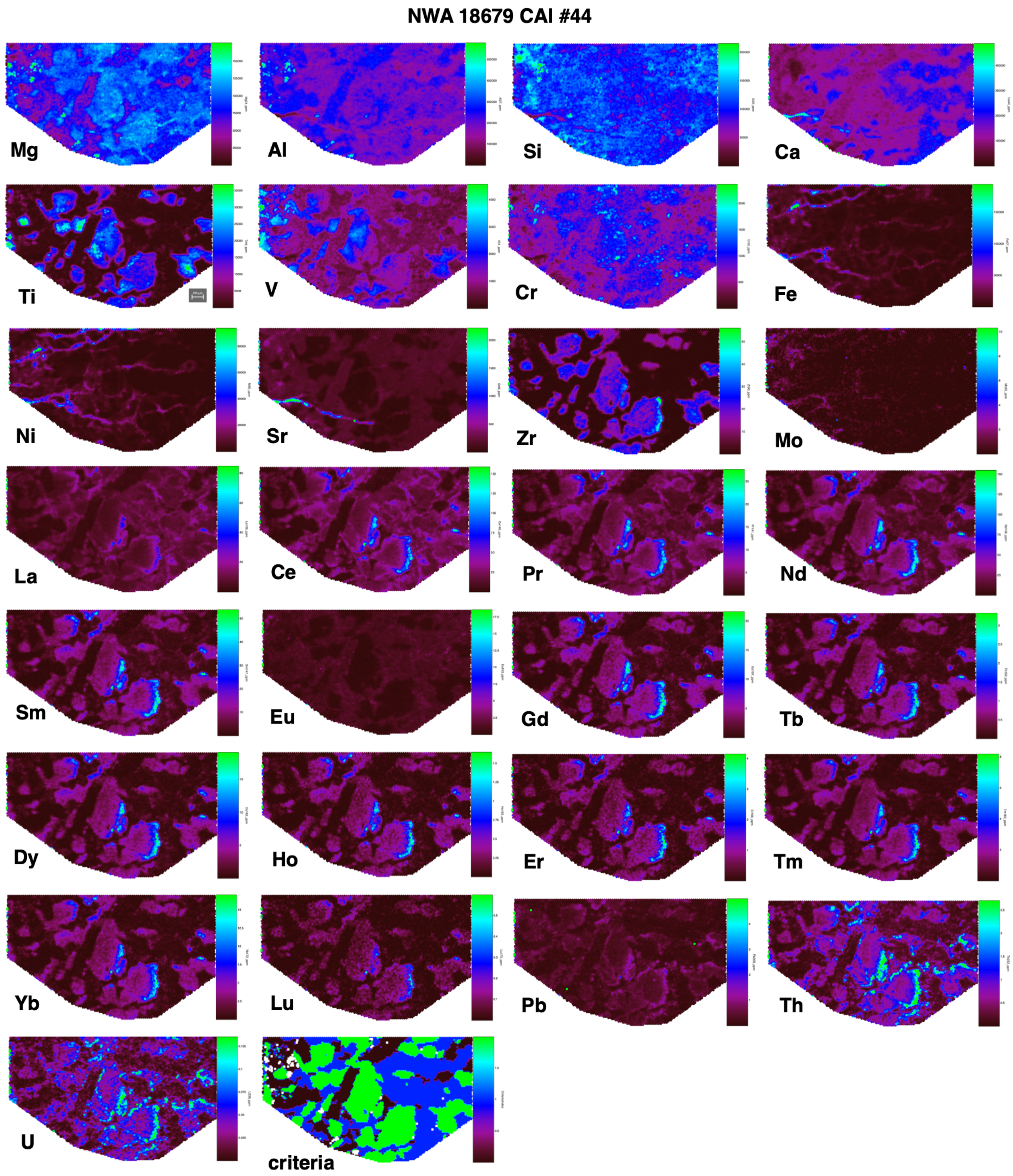

*(Fig. S3, continued)*

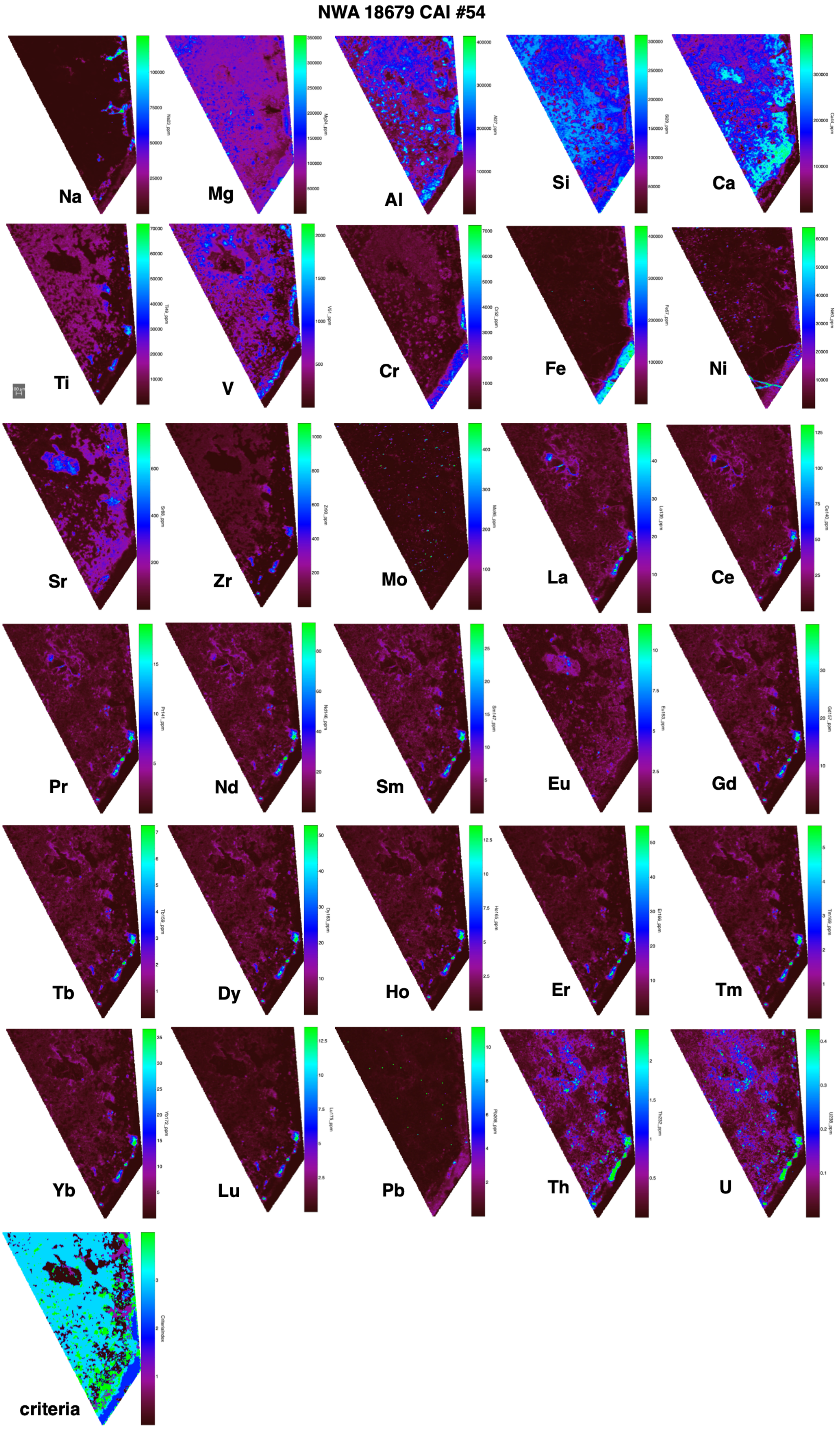

*(Fig. S4, continued)*

NWA 18679 CAI #55

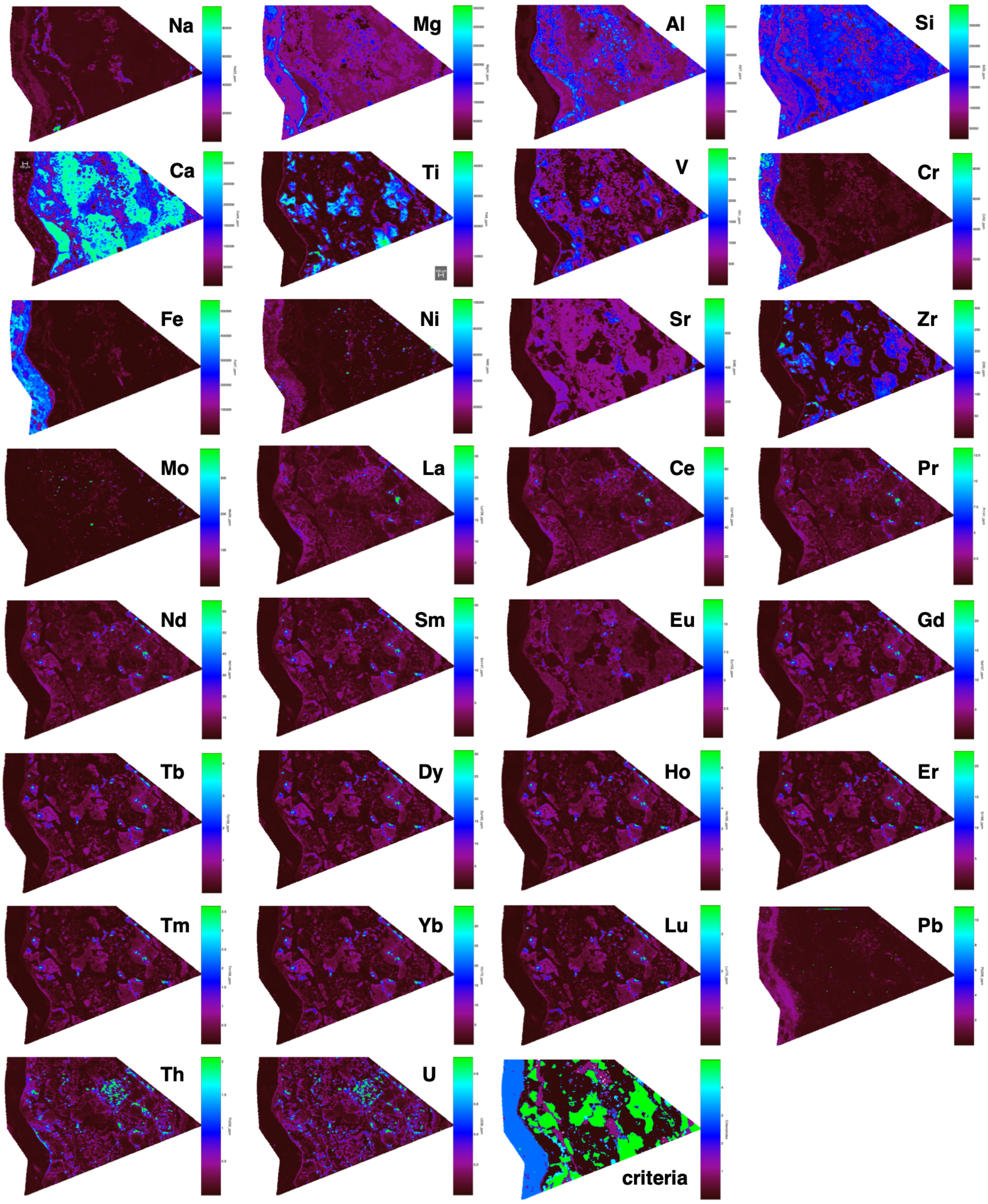

*(Fig. S5, continued)*

Leoville CAI L1B-1

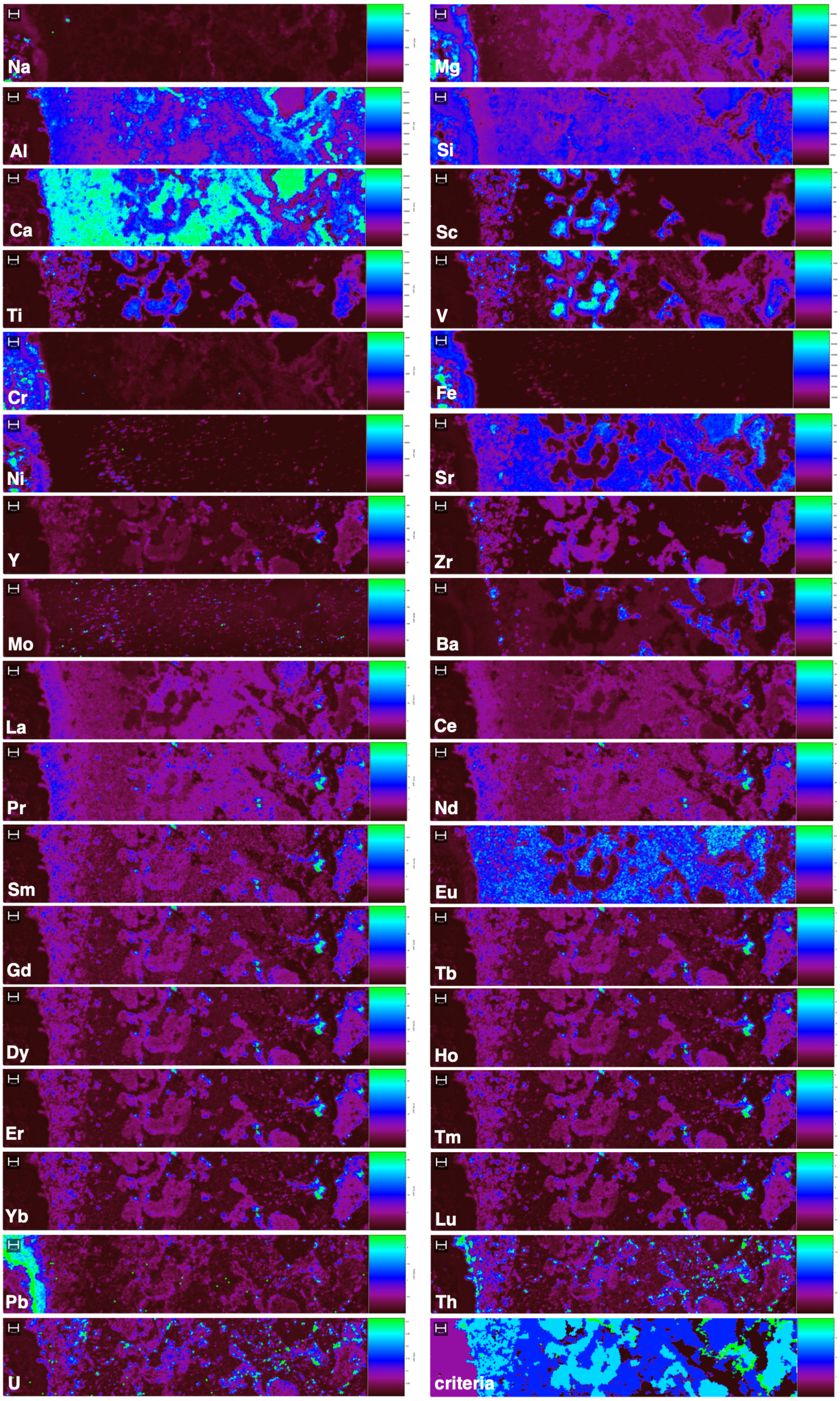

*(Fig. S6, continued)*

**Leoville CAI L1B-1: relict CAI**

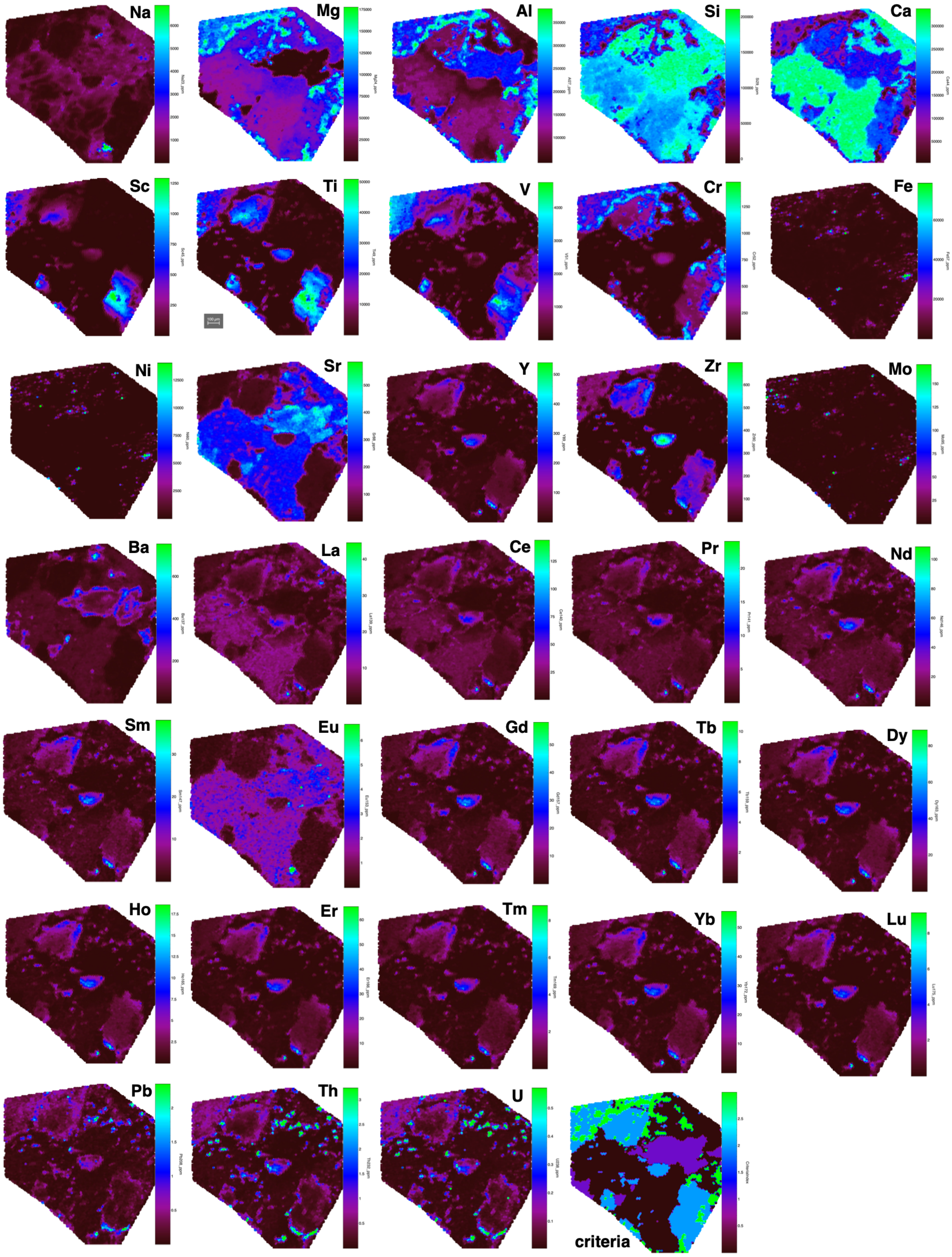

*(Fig. S7, continued)*

Leoville CAI L1B-2

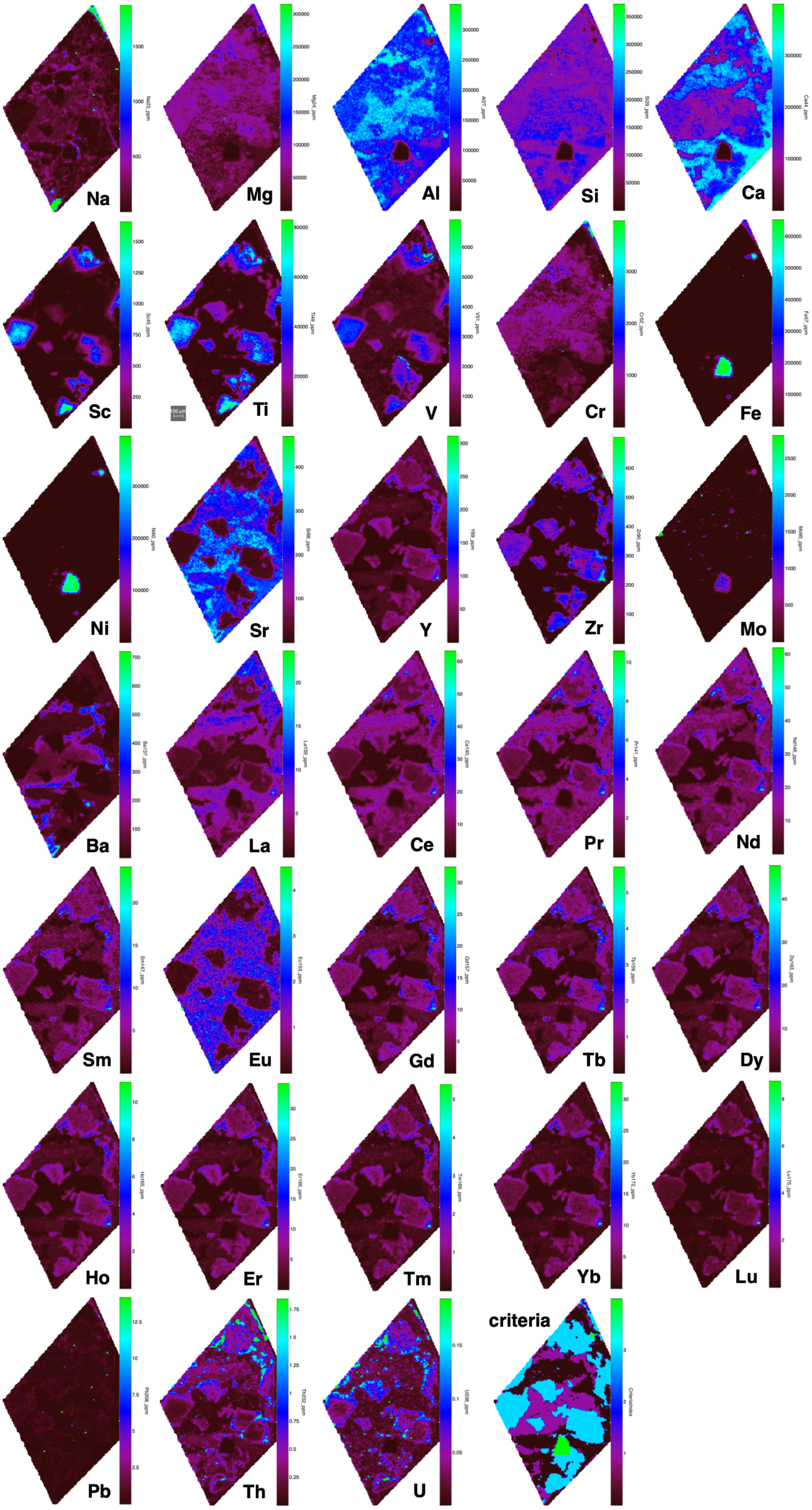

*(Fig. S8, continued)*

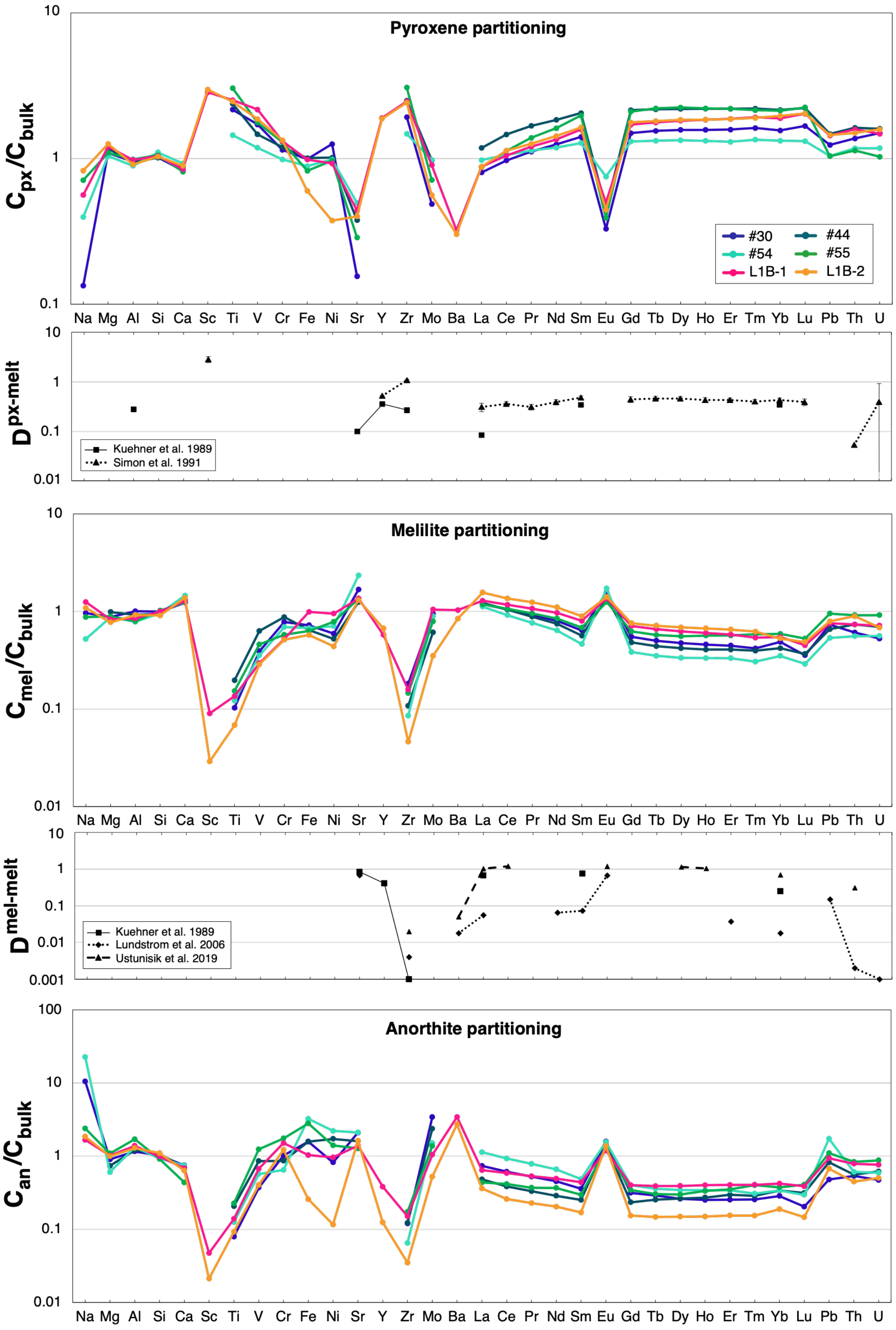

**Figure S9:** Mineral partitioning behaviour. The average elemental concentrations of Al,Ti-pyroxene, melilite and anorthite are normalized to the bulk concentrations of their host inclusion to show consistent mineral partitioning behaviour across the coarse-grained CAIs. This partitioning behaviour is broadly consistent with published mineral-melt partition coefficients ($D^{min\text{-}melt}$), shown in black below the panels of pyroxene and melilite.

**Supplementary Methods: Cm-U decay modelling details**

In order to calculate the variable impact of Cm-U partitioning on the U isotopic compositions of CAI minerals, we need to calculate the extent of Cm/U fractionation in each phase (i.e. $^{247}Cm/^{238}U$ at $t_0$). As no long-lived isotopes of Cm exist today, we use Nd and Th as proxies for the geochemical behaviour of Cm (Brennecka et al., 2010; Tissot et al., 2016). This means that we assume no fractionation occurs between Cm and Nd or Cm and Th during CAI-forming processes, which corresponds mainly to igneous processes for coarse-grained CAIs thought to crystallize from a melt. The $^{247}Cm/^{144}Nd$ and $^{247}Cm/^{232}Th$ ratios at $t_0$ are therefore considered invariant (Eq. 1, 2). Both Brennecka et al. (2010) and Tissot et al. (2016) report estimates for these ratios based on their correlation with U isotopic compositions of bulk CAIs (Table S5). The values of Brennecka et al. (2010) are slightly higher than Tissot et al. (2016), resulting in higher $^{247}Cm/^{238}U$ at $t_0$ and a stronger impact of $^{247}Cm$ decay on the U isotopic composition. However, we consider the values of Tissot et al. (2016) more accurate as they are based on measurements of a U-depleted CAI ('Curious Marie') that records an anomalous U isotopic composition that cannot be explained by kinetic isotope fractionation. This inclusion provides definitive evidence that short-lived radionuclide $^{247}Cm$ was live in the early Solar System.

To determine the $^{247}Cm/^{238}U$ at $t_0$, we also calculate the $^{144}Nd/^{238}U$ and $^{232}Th/^{238}U$ ratios in the phase at $t_0$ from the $^{144}Nd/^{238}U$ and $^{232}Th/^{238}U$ atomic ratios today where we account for $^{238}U$ and $^{232}Th$ decay and assume a formation age of 4567.30 ± 0.16 Ma (Eq. S1, S2; decay constants in Table S5). There is no need to correct for $^{144}Nd$ decay as it negligible over the timespan considered here. The $^{144}Nd/^{238}U$ and $^{232}Th/^{238}U$ atomic ratios today in the phase are calculated from Nd, Th and U concentrations acquired by fs-LA-ICP-MS.

$$\left(\frac{^{247}Cm}{^{238}U}\right)_{phase,\,t_0} = \left(\frac{^{247}Cm}{^{144}Nd}\right)_{t_0} \times \left(\frac{^{144}Nd}{^{238}U}\right)_{phase,\,t} \times e^{-\lambda_{238}t} \qquad \text{(S1)}$$

$$\left(\frac{^{247}Cm}{^{238}U}\right)_{phase,\,t_0} = \left(\frac{^{247}Cm}{^{232}Th}\right)_{t_0} \times \left(\frac{^{232}Th}{^{238}U}\right)_{phase,\,t} \times e^{(\lambda_{232}-\lambda_{238})t} \qquad \text{(S2)}$$

Without $^{247}Cm$ decay as a production source of $^{235}U$, the U isotopic composition today is determined from its initial composition at any given time:

$$\left(\frac{^{235}U}{^{238}U}\right)_{phase,\,t} = \left(\frac{^{235}U}{^{238}U}\right)_{phase,\,t_0} \times e^{-(\lambda_{235}-\lambda_{238})t} \qquad \text{(S3)}$$

To account for the ingrowth of $^{235}U$ through $^{247}Cm$ decay, a term is added to equation S3 that corresponds to the Bateman equation, which accounts for a decaying daughter:

$$\left(\frac{^{235}U}{^{238}U}\right)_{phase,\,t} = \left(\frac{^{235}U}{^{238}U}\right)_{phase,\,t_0} \times e^{-(\lambda_{235}-\lambda_{238})t} + \left(\frac{^{247}Cm}{^{238}U}\right)_{phase,\,t_0} \times \left(\frac{\lambda_{247}}{\lambda_{235}-\lambda_{247}}\right) \times \left(e^{-(\lambda_{247}-\lambda_{238})t} - e^{-(\lambda_{235}-\lambda_{238})t}\right) \text{(S4)}$$

The initial U isotopic composition of CAIs is likely set by kinetic isotopic fractionation during condensation and evaporation processes which means there is no one correct initial value for all CAIs (Amelin et al., 2010; Connelly et al., 2012; Tissot et al., 2016). Although the final U isotopic composition is dependent on this initial value, in this study we focus on the changes to the composition caused by variable Cm-U partitioning into CAI minerals. As such, instead of an initial U isotopic composition that evolves, we replace this term in equation 4 with today's $^{238}U/^{235}U$ ratio of 137.844 in equation S5 (i.e. U isotopic composition of standard CRM-145; Condon et al., 2010). Varying this value has no effect on the age offsets obtained.

$$\left(\frac{^{235}U}{^{238}U}\right)_{phase,\,t} = \left(\frac{1}{137.844}\right) + \left(\frac{^{247}Cm}{^{238}U}\right)_{phase,\,t_0} \times \left(\frac{\lambda_{247}}{\lambda_{235}-\lambda_{247}}\right) \times \left(e^{-(\lambda_{247}-\lambda_{238})t} - e^{-(\lambda_{235}-\lambda_{238})t}\right) \quad \text{(S5)}$$

We report the model $^{235}U/^{238}U$ compositions in mineral phases in the $\varepsilon^{235}U$ notation, which represents the parts per ten thousand deviations of the sample or mineral phase relative to the U isotopic composition of the standard CRM-145 (Eq. S6; Table S2). In equation S7, we replace the $^{235}U/^{238}U$ composition of the phase at t by right-hand side expression of equation S5, so we can directly calculate this value from $^{247}Cm/^{238}U$ at $t_0$.

$$\varepsilon^{235}U = \left(\left(\left(\frac{^{235}U}{^{238}U}\right)_{phase,t} \times 137.844\right) - 1\right) \times 10000 \quad \text{(S6)}$$

$$\varepsilon^{235}U = \left(\left(\left(\frac{^{247}Cm}{^{238}U}\right)_{phase,\,t_0} \times \left(\frac{\lambda_{247}}{\lambda_{235}-\lambda_{247}}\right) \times \left(e^{-(\lambda_{247}-\lambda_{238})t} - e^{-(\lambda_{235}-\lambda_{238})t}\right) \times 137.844\right) - 1\right) \times 10000 \quad \text{(S7)}$$

We estimate the uncertainties on the $^{247}Cm/^{238}U$ values and resulting $\varepsilon^{235}U$ values through Monte Carlo error propagation ($n=10^5$) because these calculations are non-linear and complex. These uncertainties account for the uncertainties on the initial $^{247}Cm/^{144}Nd$ and $^{247}Cm/^{232}Th$ ratios from Tissot et al. (2016), the 2SE uncertainties on concentration data from this study, the uncertainty on the CAI formation age (Connelly et al., 2012) and the uncertainties on the decay constants (Table S5). The reported model values represent the mean of the Monte Carlo outputs and the uncertainties represent the standard deviation of the outputs (Table S2).

**Table S1**: Major- and trace-elemental abundances of CAIs
**Table S2**: Nd and Th as proxies for Cm to calculate mineral-specific U isotope compositions
**Table S3**: Mineral partitioning behaviour
**Table S4**: Modelled age offsets in mismatching scenarios
*(Table S1 to S4 in Supplementary Excel file)*

| Parameter | Values | References/Notes |
|---|---|---|
| $\left(\frac{^{247}Cm}{^{144}Nd}\right)_{t_0}$ | $(2.06 \pm 0.47) \times 10^{-6}$ | Tissot et al., 2016 - ESS |
| | $(3.3 \pm 0.6) \times 10^{-6}$ | Brennecka et al., 2010 - ESS |
| $\left(\frac{^{247}Cm}{^{232}Th}\right)_{t_0}$ | $(9.7 \pm 2.2) \times 10^{-6}$ | Tissot et al., 2016 - ESS |
| | $(1.5 \pm 0.3) \times 10^{-5}$ | Brennecka et al., 2010 - ESS |
| $\lambda_{235}$ | $(9.85 \pm 0.01) \times 10^{-10}$ yr$^{-1}$ | $^{235}$U decay constant |
| $\lambda_{238}$ | $(1.551 \pm 0.003) \times 10^{-10}$ yr$^{-1}$ | $^{238}$U decay constant |
| $\lambda_{247}$ | $(4.45 \pm 0.14) \times 10^{-8}$ yr$^{-1}$ | $^{247}$Cm decay constant |
| $\lambda_{232}$ | $(4.95 \pm 0.02) \times 10^{-11}$ yr$^{-1}$ | $^{232}$Th decay constant |

**Table S5**: Parameter values used for $^{247}$Cm-$^{235}$U decay modelling of mineral phases in CAIs. Model results are reported using the values of Tissot et al. (2016).